\documentclass[usenatbib]{mnras}
\usepackage[T1]{fontenc}
\usepackage{ae,aecompl}
\usepackage{ulem} 
\usepackage{color} \usepackage{textcomp}
\usepackage{enumerate} \usepackage{amssymb} \usepackage{txfonts}
\usepackage{url}
\usepackage{multirow}
\usepackage{diagbox}
\usepackage{blindtext}
\usepackage{graphicx}	
\usepackage{natbib, epsfig}
\title[Comparison of two full-spectrum fitting algorithms]{Recovering stellar population parameters via two full-spectrum fitting algorithms in the absence of model uncertainties} 
\author[J. Q. Ge et al.]{ 
Junqiang Ge,$^1$\thanks{E-mail: jqge@nao.cas.cn}
Renbin Yan,$^2$ 
Michele Cappellari,$^3$ 
Shude Mao,$^{4,1,5}$ 
Hongyu Li,$^1$ 
Youjun Lu$^1$\\
$^{1}$National Astronomical Observatories, Chinese Academy
  of Sciences, 20 Datun Road, Beijing 100020, China\\
$^{2}$Department of Physics and Astronomy, University of Kentucky, 
               505 Rose Street, Lexington, KY 40506, USA\\
$^{3}$Sub-Department of Astrophysics, Department of Physics, 
               University of Oxford, Denys Wilkinson Building, Keble Road, 
              Oxford, OX1 3RH, UK\\
$^{4}$Physics Department and Tsinghua Center for Astrophysics,
               Tsinghua University, Beijing, 100084, China\\
$^{5}$Jodrell Bank Centre for Astrophysics,
               Alan Turing Building, The University of Manchester,
               Manchester M13 9PL, UK
}
\date{Accepted XXX. Received YYY; in original form ZZZ}
\pubyear{2018}

\begin{document}
\pagerange{\pageref{firstpage}--\pageref{lastpage}}
\maketitle

\begin{abstract}
Using mock spectra based on Vazdekis/MILES library fitted within the wavelength region 3600-7350\AA, 
we analyze the bias and scatter 
on the resulting physical parameters induced by the choice of fitting algorithms and 
observational uncertainties, but avoid effects of those model uncertainties. We consider two 
full-spectrum fitting codes: pPXF and STARLIGHT, in fitting for stellar population age, 
metallicity, mass-to-light ratio, and dust extinction. 
With pPXF we find that both the bias $\mu$ in the population parameters and the 
scatter $\sigma$ in the recovered logarithmic values follows the expected trend 
$\mu\propto\sigma\propto1/(\rm S/N)$. The bias increases for younger ages and systematically 
makes recovered ages older, $M_*/L_r$ larger and  metallicities lower than the true values. 
For reference, at S/N=30, and for the worst case ($t=10^8$yr), the bias is 0.06 dex in $M_*/L_r$, 
0.03 dex in both age and [M/H]. There is no significant dependence on either E(B-V) or the 
shape of the error spectrum. Moreover, the results are consistent for both our 1-SSP 
and 2-SSP tests.
With the STARLIGHT algorithm, we find trends similar to pPXF, when the input E(B-V)$<0.2$ mag. 
However, with larger input E(B-V), the biases of the output parameter do not converge 
to zero even at the highest S/N and are strongly affected by the shape of the error 
spectra. This effect is particularly dramatic for youngest age ($t=10^8$yr), for which all 
population parameters can be strongly different from the input values, with significantly 
underestimated dust extinction and [M/H], and larger ages and $M_*/L_r$. Results degrade when 
moving from our 1-SSP to the 2-SSP tests. The STARLIGHT convergence to the true values can 
be improved by increasing Markov Chains and annealing loops to the ``slow mode". 
For the same input spectrum, pPXF is about two order of magnitudes faster than STARLIGHT's 
``default mode'' and about three order of magnitude faster than STARLIGHT's ``slow mode''.
\end{abstract}

\section{Introduction} 
\label{sec:introduction}
One important way of understanding galaxy formation and evolution is to constrain
stellar population properties with stellar population synthesis.
However, due to the degeneracy of different parameters, such as age,
metallicity, dust extinction, and initial mass function (IMF), and due to uncertainties in
stellar evolution model and stellar spectral library, results from stellar population
synthesis may vary with different algorithms and stellar population models 
\citep[see reviews in ][]{walcher2011, conroy2013}.

Stellar evolution models have been studied for many decades, initially from studying stellar populations 
at a certain age and metallicity \citep[e.g.][]{tinsley1968, ssb1973, tg1976, bruzual1983}, then
improved to modelling stellar evolutions at the whole age and metallicity parameter spaces \citep[][]{cb1991, bc1993, 
bcf1994, worthey1994, frv1997, maraston1998, leitherer1999, vazdekis1999, walcher2011}.
With many efforts dedicated to stellar evolution theory, there are several well developed popular models, 
such as Padova \citep[][]{bertelli1994,girardi2000, marigo2008}, 
BaSTI \citep[][]{pietrinferni2004, cordier2007}, and Geneva \citep[][]{schaller1992, mm2000}.

Accompanying the development of stellar evolution models, there are also two kinds of stellar population synthesis
models: 1) empirical population synthesis method, 
and 2) evolutionary population synthesis method. The first method tries to reproduce a galaxy spectrum
by means of a linear combination of individual stellar spectra with various types taken
from a comprehensive library \citep{wood1966, mv1968, st1971, faber1972, 
bica1988, pelat1997, cid2001, moultaka2004}.
The second method was developed 
at almost the same time by comparing galaxy data with synthesized stellar population spectrum based on 
stellar evolution theory, and with the IMF, star formation and chemical histories as main
adjustable parameters \citep{tinsley1978, bruzual1983, ay1987, gr1987, buzzoni1989, 
bc1993, bcf1994, worthey1994, lh1995, frv1997, maraston1998, vazdekis1999, schulz2002}.

Combining with assumed IMFs and empirical stellar spectral libraries, e.g. \cite{gs1983}, 
\cite{pickles1998}, \cite{jones1999}, ELODIE \citep{ps2001}, STELIB \citep{leborgne2003}, 
Indo-US \citep{valdes2004}, NGSL \citep{gregg2006, hl2011}, 
MILES \citep{sb2006}, IRTF \citep{rayner2009}, 
and the X-shooter library \citep{chen2011}, many stellar population models are now available 
for full-spectrum fitting, such as BC03 \citep{bc2003}, 
FSPS \citep[][]{cgw2009}, Vazdekis/MILES \citep{vazdekis2010}, and M11 \citep{ms2011}.

With the improved stellar population models, faster computing capabilities, and
the availability of modern spectroscopic galaxy surveys, such as the Sloan Digital Sky Survey \citep[SDSS;][]{york2000},
the spectral 
population analysis has transitioned from the modelling based on a few observables, such as colours, absorption-line
equivalent width or spectral indices \citep[e.g.][]{wood1966, faber1972, worthey1994, kauffmann2003},
to the more precise pixel-by-pixel full-spectrum fitting
that exploits the full spectral information \citep[e.g.][]{CappellariEmsellem2004, cid2005, tojeiro2007, koleva2009}.
The algebraic shortage, which is caused by the number of unknowns larger than observables
when fitting with several colours or spectral indices,
is no longer a problem in the full-spectrum fitting era.
Those well-calibrated synthetic spectra at high spectral resolution have dramatically 
improved the possibility of using full spectrum fitting, 
instead of line indices, 
to study stellar populations. Now there are a number of spectral fitting codes, 
such as pPXF \citep[][]{CappellariEmsellem2004, Cappellari2017}, 
STARLIGHT \citep[][]{cid2005}, STECKMAP \citep[][]{ocvirk2006}, VESPA \citep[][]{tojeiro2007}, 
ULYSS \citep[][]{koleva2009}, Fit3D \citep[][]{sanchez2011}, FIREFLY \citep[][]{wilkinson2015} 
and FADO \citep[][]{gp2017}, written with different algorithms.

Given all these advances of stellar population analysis methods and their wide spread applications, 
it is essential to understand their systematic biases and uncertainties, especially those from
full-spectrum fitting methods. Only by understanding its limitations, can we truly understand its power and continue to improve it. 

There have been many works focusing on the reliability test for constraining stellar population using 
broadband spectral energy distributions (SED) \citep[e.g.][]{pdf2001, shapley2001, wuyts2009, muzzin2009, lee2009}. 
These works found that the stellar mass measurement tends to be more reliable than 
others, such as stellar age, metallicity, dust extinction ($A_V$ or E(B-V)), and star formation rate (SFR).
Considering that young stars can outshine older ones, for the one-component SED fitting, 
the stellar mass can be underestimated when both young and old stars exist \citep[e.g. 19\%-25\%,][]{lee2009}.
Furthermore, the stellar mass-to-light ratios ($M_*/L$) obtained from SED fitting with simple stellar population (SSP) models or single-age models 
are lower than those with complex SFH cases \citep{pdf2001, shapley2001, tfd2008, gf2010, pmt2012}.
If a galaxy contains only young stars and one allows for both young and old in the fit, 
there will be a tendency to over-estimate its mass, due to noise allowing a small amount of old models, 
and the biased $M_*/L$ varies with stellar ages \citep{gb2009}.
Besides the SFH effects on the stellar population analysis, 
dust extinction is another important cause for bias, which can underestimates the stellar mass by 40\% \citep{zcr2009}.

There are already some tests based on full-spectrum fitting method, such as \cite{cid2005} and 
\cite{ocvirk2006}. \cite{cid2005} have checked the recovery probability of the STARLIGHT code 
based on mock spectra with an assumed SFH and found that the output can recover the input well but with a large 
scatter, which is mainly due to uncertainties introduced by error spectrum and low signal-to-noise ratio.
\cite{ocvirk2006} checked their STECMAP code and found that two starbursts can be distinguished well 
only if they have an age separation larger than $\sim$0.8 dex with high spectral resolution (R=10,000)
and S/N (=100). There is also an extinction measurement bias for different methods as shown in Figure 8 of 
\cite{cid2005}. STARLIGHT tends to give lower dust extinction than the MPA/JHU data products.
In \cite{li2017}, we applied the pPXF and STARLIGHT codes for the spectral fitting of MaNGA IFU data, 
and find that measured $M_*/L_r$ from the two codes are consistent when $M_*/L_r>3$, 
but a bias appears and increases with decreasing $M_*/L_r$ when $M_*/L_r<3$.

There are basically three sources that introduce the systematic biases and uncertainties in the results of a stellar population analysis.
The first one is observational uncertainties such as noise in the spectra and its dependence on wavelength, 
and systematics in flux calibration. The second source includes inaccurate model assumptions, 
such as inaccuracies in the initial mass function, stellar evolution tracks, binary evolution, 
and stellar spectra library. Biases in model assumptions cause biases directly in the results of population analyses.
The third source includes inherent degeneracies among physical parameters (e.g. age, metallicity, and extinction). 
Different fitting algorithms respond differently to these issues. The first and the second categories 
can be separately tested, but the degeneracies included in the third source are always unavoidable. 

In this paper we aim at setting a baseline for what can be recovered using spectral fitting based on the pPXF and STARLIGHT codes. 
Rather than studying a complex SFH, which allows a large range of possibilities and makes the results 
difficult to interpret, we intentionally keep the assumptions extremely simple.
For simplicity, we use simulated spectra generated by one SSP and a linear combination of 
two SSPs to check the bias and uncertainty of the full-spectrum fitting, and only test how observational uncertainties 
affect the fitting results. This can tell us which kind of S/N is needed in Voronoi 2D binning \citep{cc2003} of spectral data cubes 
and in designing future observations. By fitting the simulated spectra with the same model 
assumption they are built with, we choose to ignore biases due to inaccurate model assumptions here 
and leave it for future investigations.

We will describe the related codes of the two full-spectrum fitting algorithms
and the corresponding stellar population library in Section 2, and present the bias and scatter of different parameter 
estimations in Section 3. Those matters need attention when using STARLIGHT for spectral fitting and biases when 
applying the two codes to observations are discussed in Section 4. Finally we summarize our results in Section 5.

\section{Preparation for the test}
In this section, we briefly introduce the related algorithms of the two spectral fitting codes, selected stellar library, 
and the initial parameter setup for spectral fitting.

\subsection{Full-spectrum fitting codes}

The pPXF code \citep{CappellariEmsellem2004, Cappellari2017}, which is written in both IDL and Python programs 
(here we use the latest Python version), uses a maximum penalized likelihood 
method in the pixel space to fit the spectra, with the line-of-sight
velocity distributions (LOSVD) described by Gauss-Hermite parameterization. It uses a non-negative 
least-squares (NNLS) solver \citep{LawsonHanson1974} to fit for the spectral weights, 
embedded into a novel Levenberg-Marquardt solver 
adapted for bound constraints to fit the nonlinear parameters, 
describing the galaxy kinematics and reddening. Most applications were for stellar kinematics, 
but there were a number of applications for stellar population analyses 
\citep[][]{cappellari2012, onodera2012, morelli2013, morelli2015, mcdermid2015, sc2015, li2017}.
The current Python/IDL versions of pPXF can also fit gas components at the same time.

The STARLIGHT code \citep{cid2005} treats all parameters as nonlinear and determines their optimal values with a 
simulated annealing plus Metropolis scheme to search for the minimum $\chi^2$.
Through the Metropolis algorithm, this scheme gradually focuses on the most likely region in the parameter space by avoiding
trapping in local minima. In some applications, 
the Markov chain generated by the Metropolis algorithm can remain trapped in a local minimum 
and the global convergence may not be reached \citep[e.g.][]{martino2012}.
After combining with simulated annealing to avoid trapping, \cite{cid2005} have checked the consistency 
between input vs. output results, and found that the dust extinction had a clear difference with that from the 
MPA/JHU group. To have a thorough idea on the algorithm bias, more detailed analyses are required, which
we perform in this work.

\subsection{Stellar population library}

We adopt the Vazdekis/MILES simple stellar population library \citep{vazdekis2010} for full-spectrum fitting, 
by assuming a Salpeter IMF \citep{salpeter1955} with stellar mass range $[0.1, 100]M_{\odot}$, Padova 2000 
stellar evolution model \citep{girardi2000}, and MILES stellar spectral library \citep{sb2006}. 
We select a subset of 25 logarithmically-spaced, equally-sampled ages between 0.0631 and 15.8489 Gyr inclusive
and 6 metallicities ([M/H]=$-1.71, -1.31, -0.71, -0.4, 0.0, 0.22$).

\begin{figure}
\centering
\includegraphics[angle=0.0,scale=0.6,origin=lb]{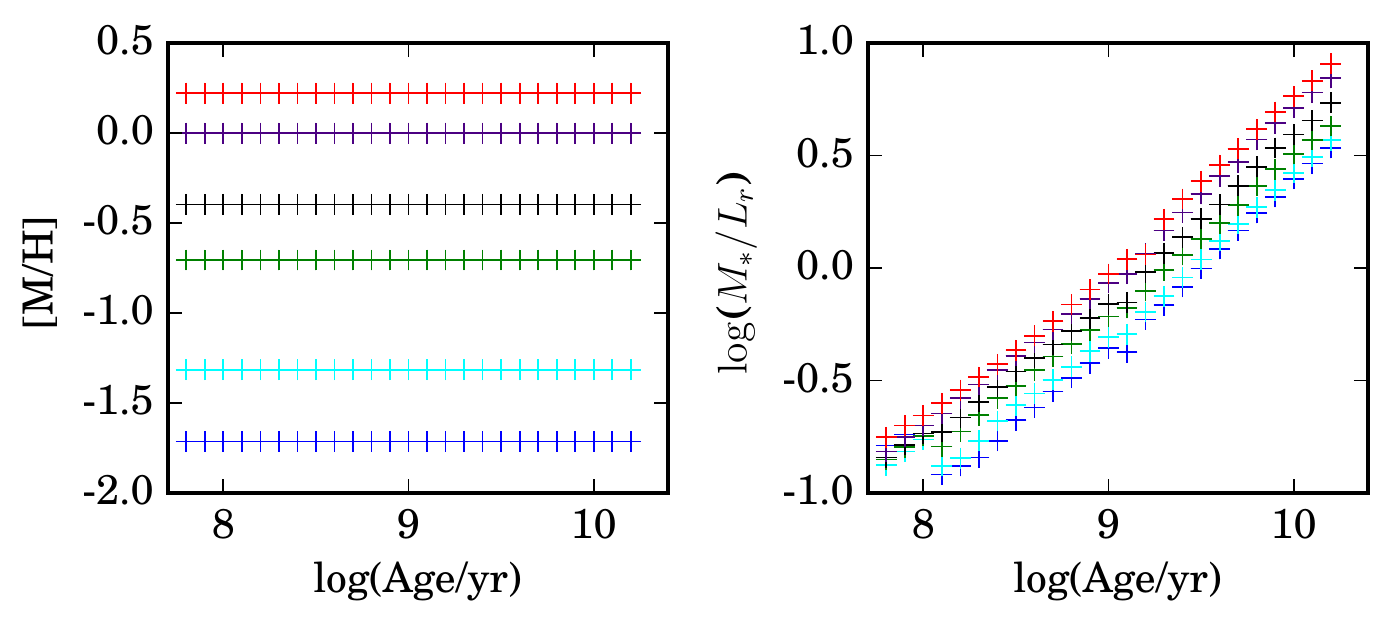}
\caption{The stellar age, metallicity, and mass-to-light ratio distributions of Vazdekis/MILES stellar population
library, generated by assuming a Salpeter IMF and Padova 2000 stellar evolution tracks.
Left: Stellar age and m
etallicity distribution;
Right: $M_*/L_r$ along stellar evolution. SSPs with different metallicities are shown in different colors, 
$\rm [M/H]$= $-1.71$ (blue), $-1.31$ (cyan), $-0.71$ (green), $-0.4$ (black), 0.0 (violet), 0.22 (red).
}
\label{stel_lib}
\end{figure}

Figure\ref{stel_lib} shows the parameter distributions of stellar age vs. metallicity ($left$),
and stellar age vs. $M_*/L_r$ ($right$). 
SSPs with different metallicities are shown in different colors.
The corresponding $M_*/L_r$ are the tabulated values provided by MILES team.
SSPs with the same age but higher metallicities usually have lower mass loss rate and higher
$M_*/L_r$ ($right~ panel$) than that with lower metallicities.

Since the MILES stellar spectral library have a broader fundamental parameter coverage than others \citep{vazdekis2010},
adopting the current Vazdekis/MILES stellar population library becomes a logical choice for checking 
full-spectrum fitting algorithms.

\subsection{Initial parameter setup}
We generate the mock spectra based on the Vazdekis/MILES library with FWHM=2.54\AA, then use the same library for spectral fitting, 
to avoid biases due to incorrect model assumptions on IMFs, stellar evolution models, stellar spectral libraries, 
and dust extinction curves.
The independent input variables include the dust extinction, age, and metallicity.
The spectral fitting range of mock data is 3600-7350\AA, which is shorter by $\sim$50\AA~ than the models at both the blue and red end to make sure
the model SSPs have larger spectral coverage than mock spectra.

As the SDSS IV/MaNGA survey \citep{Bundy2015} is planning to observe 10,000 galaxies, which has finished one half and is 
currently the largest galaxy sample with IFU observations, we thus set our mock spectra to mimic MaNGA's spectral 
resolution with a line spread function with FWHM=2.76\AA~ \citep{albareti2016}. To generate a mock 
spectrum, we first take a template spectrum from the Vazdekis/MILES SSP library, and smooth the 
mock spectra from FWHM=2.54\AA~ to FWHM = 2.76\AA. All the test in our 
paper will feed log-rebinned spectra to the pPXF and linear-rebinned spectra to STARLIGHT
to fulfill the requirements of pPXF and STARLIGHT codes.
The velocity scale is then set to 69 km/s for spectra sampled in the logarithmic wavelength grid, 
and wavelength scale is set to 1\AA~ for spectra sampled in the linear wavelength grid\footnote{With this sampling, 
the linear spectra has 3750 pixels, while the log spectrum has 3057. 
This implies that the effective S/N of the linear spectrum is $\sqrt{3750/3057}\sim 11$\% larger.}.
To make our mock spectra more like observed ones, we put in an additional velocity dispersion of 100 km/s, 
on top of the MaNGA spectral resolution. During the fitting, we use the full spectral information
to check the algorithm precision without masking any emission-line regions.
The current version of pPXF code can fit both the stellar emission and gas emission together,
while the STARLIGHT code can only fit the stellar spectra. To have a direct comparison between these two codes, 
we avoid tests with the emission line fitting process.

\begin{figure*}
\centering
\includegraphics[angle=0.0,scale=0.8,origin=lb]{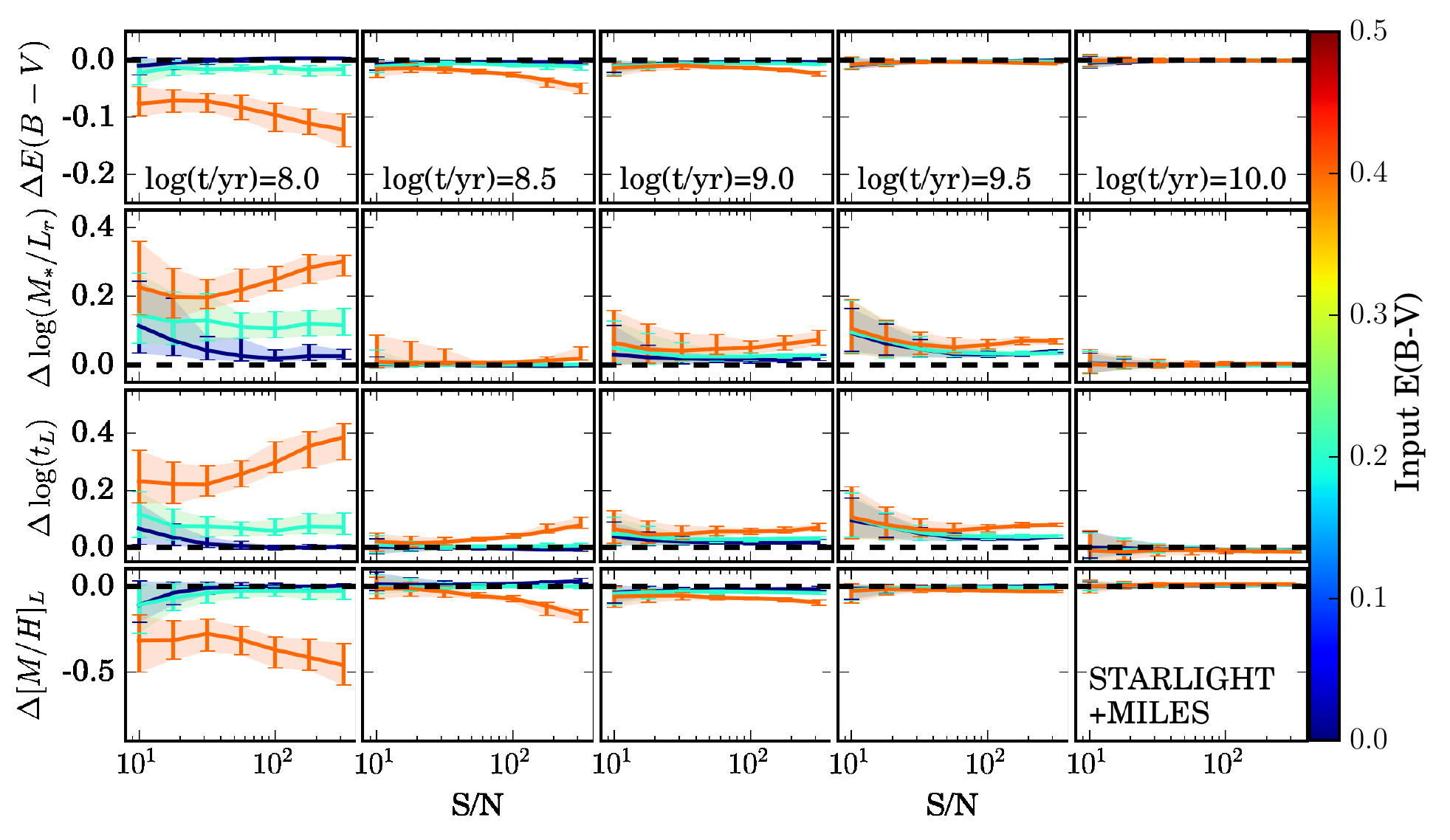}\\
\includegraphics[angle=0.0,scale=0.8,origin=lb]{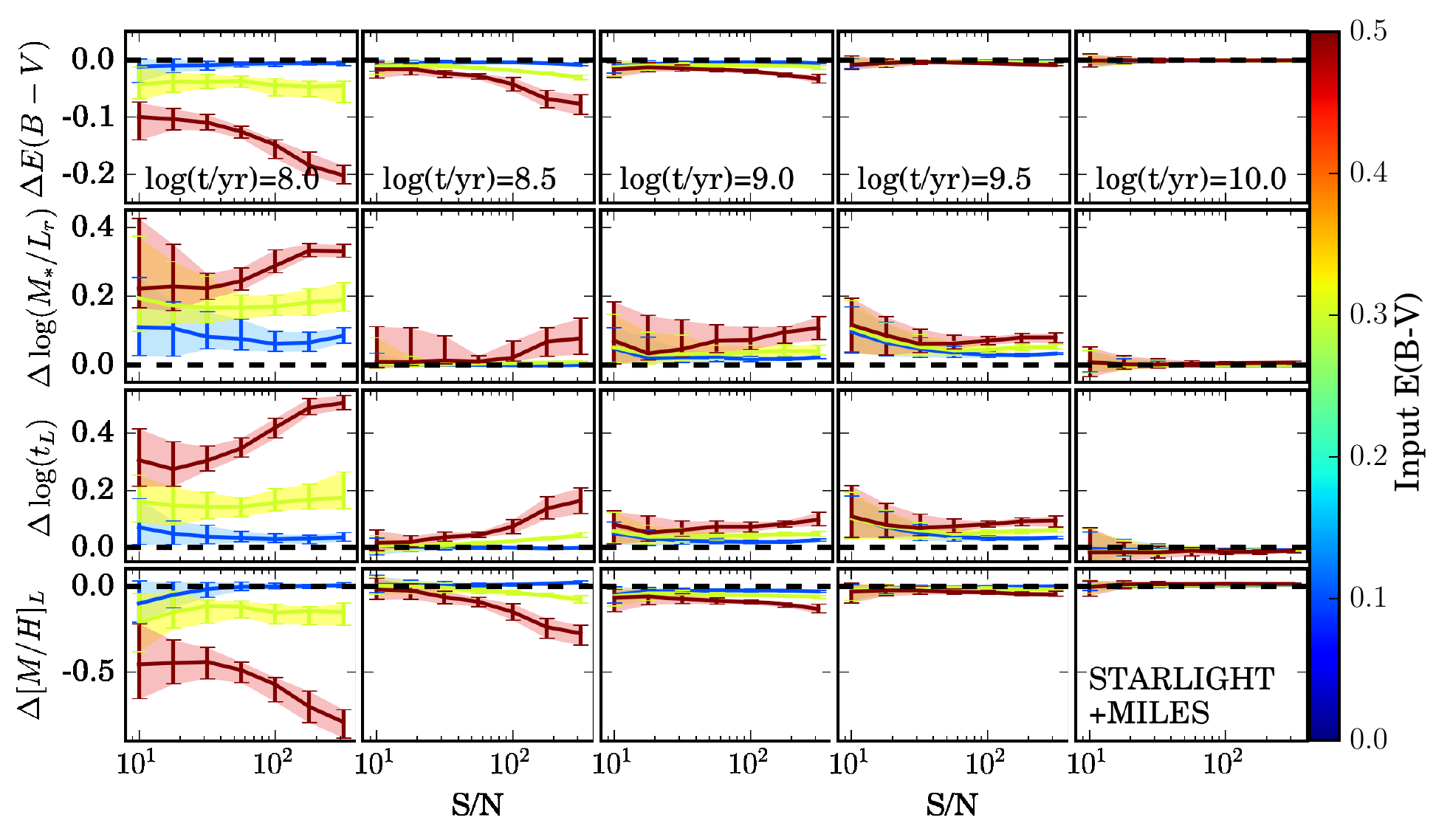}
\caption{
The bias and scatter of four stellar population parameters obtained from STARLIGHT full-spectrum 
fitting by assuming different spectral S/N. The top figure represents those results with input E(B-V)=[0, 0.2, 04], 
while the bottom one plots results with input E(B-V)=[0.1, 0.3, 0.5]. For clarity, we plot the results into two figures.
For each figure, from top row to bottom row, the parameters are $\Delta$E(B-V), $\Delta \log(M_*/L_r)$, $\Delta \log(t_L/{\rm yr})$ and 
$\Delta [M/H]_L$, respectively. From left to right column, we show the results for the mock spectra with young
 ($\log(t/{\rm yr})=$ 8.0) to old stellar ages ($\log(t/{\rm yr})=$ 10.0), and all with solar metallicity.
Lines colored from blue to red represent the increased input E(B-V)
from 0.0 to 0.5. For each line, a shaded region with a corresponding color is added to 
show the parameter scatter region between the global 
16th and 84th percentiles for input E(B-V) between 0 and 0.5.
The horizontal dashed line at each panel show the zero-bias positions along S/N. 
Each point is the median value obtained from 50 simulations. The error bars 
indicate the 16th and 84th percentiles.
}
\label{st_snr}
\end{figure*}

For dust extinction curves, in all the simulations, 
we adopt the CAL \citep{calzetti2000} dust extinction curve. Considering that those oldest elliptical
galaxies have nearly no dust extinction, while those younger starburst galaxies or those with
dust lane have higher extinction, we set the input E(B-V) from 0.0 to 0.5, 
which covers the typical E(B-V) ranges in the local universe. When applying the pPXF code for spectral fitting,
no input parameter is modified except we limit E(B-V) fitting range to [-0.125, 1], which corresponds
to $A_V$ range [-0.5, 4] with $R_V=4.05$ \citep{calzetti2000}. 
The E(B-V) should be always non-negative. However, in the case of low S/N and input E(B-V)=0, 
due to the large flux uncertainty, not only the spectral absorption lines but also the continuum are highly contaminated 
and hence can be re-shaped. 
Imposing strictly E(B-V)$\ge$0 can introduce artificially too small scatters.
In order to make sure the least $\chi^2$ is searched, negative E(B-V) is then allowed.

For the STARLIGHT code, given the range of possible configuration settings, we decided to adopt an 
identical configuration as used in the state-of-the-art analysis of the CALIFA dataset 
by \cite{deAmorim2017} and de Amorim (priv. comm.). The STARLIGHT setup was used to produce 
the CALIFA stellar population parameters publicly released in the PyCASSO database. Specifically, we adopt
the default setup from config file ``StCv04.C11.config" (included in the download package
\footnote{\url{http://www.starlight.ufsc.br/node/3}}), 
but with the normalization window changed to that in \cite{deAmorim2017}:
$\rm l\_norm=5635$\AA, $\rm llow\_norm=5590$\AA, $\rm lupp\_norm=5680$\AA, and $A_V$ fitting range to [-0.5, 4]
to allow negative $A_V$ and large enough parameter space for fitting as described in \cite{cid2005}.

\section{Full-spectrum fitting tests}

\begin{figure*}
\centering
\includegraphics[angle=0.0,scale=0.8,origin=lb]{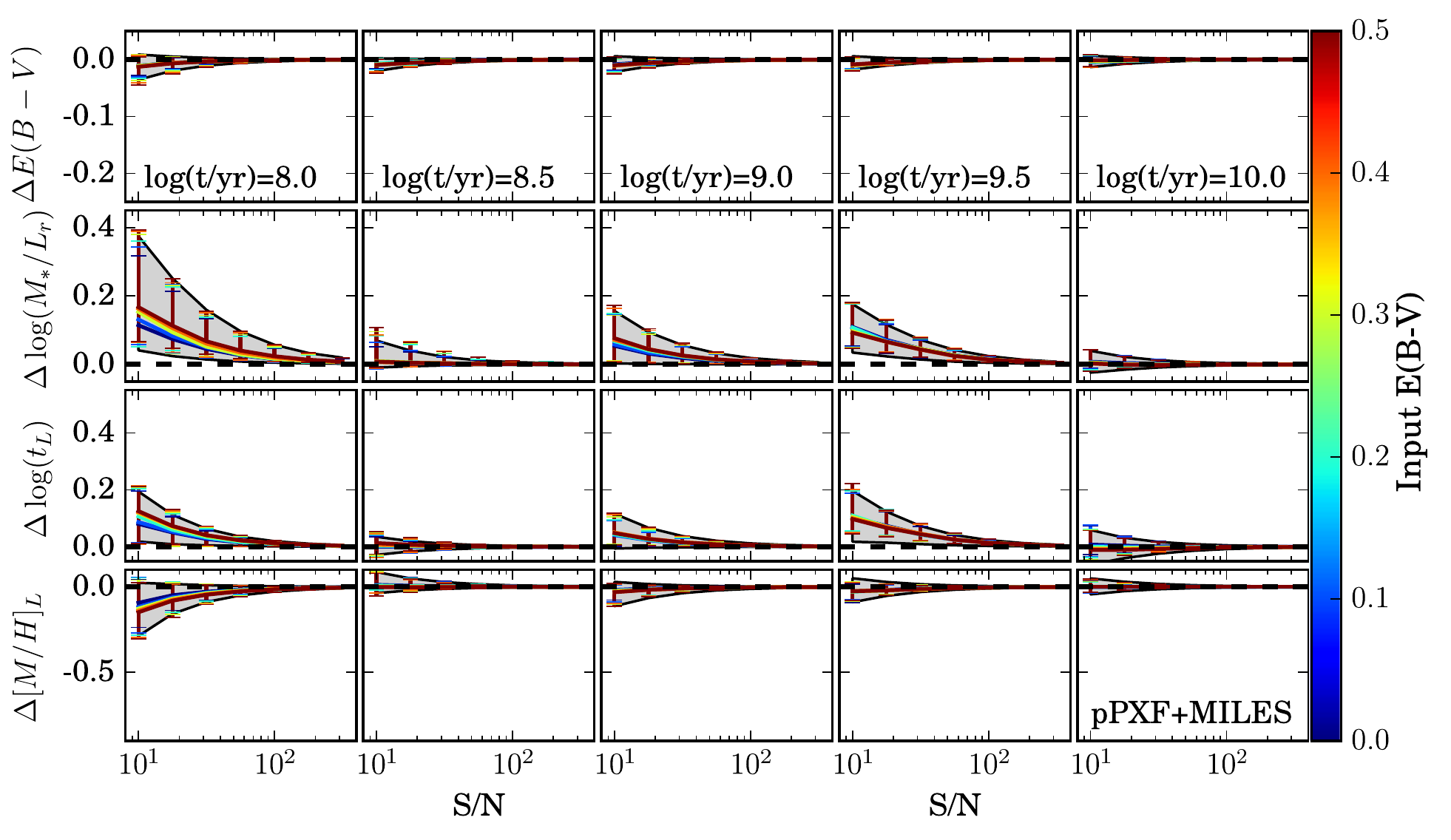}
\caption{
The bias and scatter of four stellar population parameters obtained from the pPXF full-spectrum 
fitting by assuming different spectral S/N's.
Those lines, colors, and the axis coverages in each panel are the same as shown in Figure \ref{st_snr}.
Considering that the pPXF fitting is much faster than STARLIGHT, we perform 1000 simulations for each spectrum 
and found that the results from input E(B-V)$\ge 0.1$ are almost the same. Therefore, we add a grey shaded region 
to each panel to show the parameter bias regions between the global 16th and 84th percentiles for input E(B-V) between 0 and 0.5.
}
\label{ppxf_snr}
\end{figure*}
\begin{figure*}
\centering
\includegraphics[angle=0.0,scale=0.8,origin=lb]{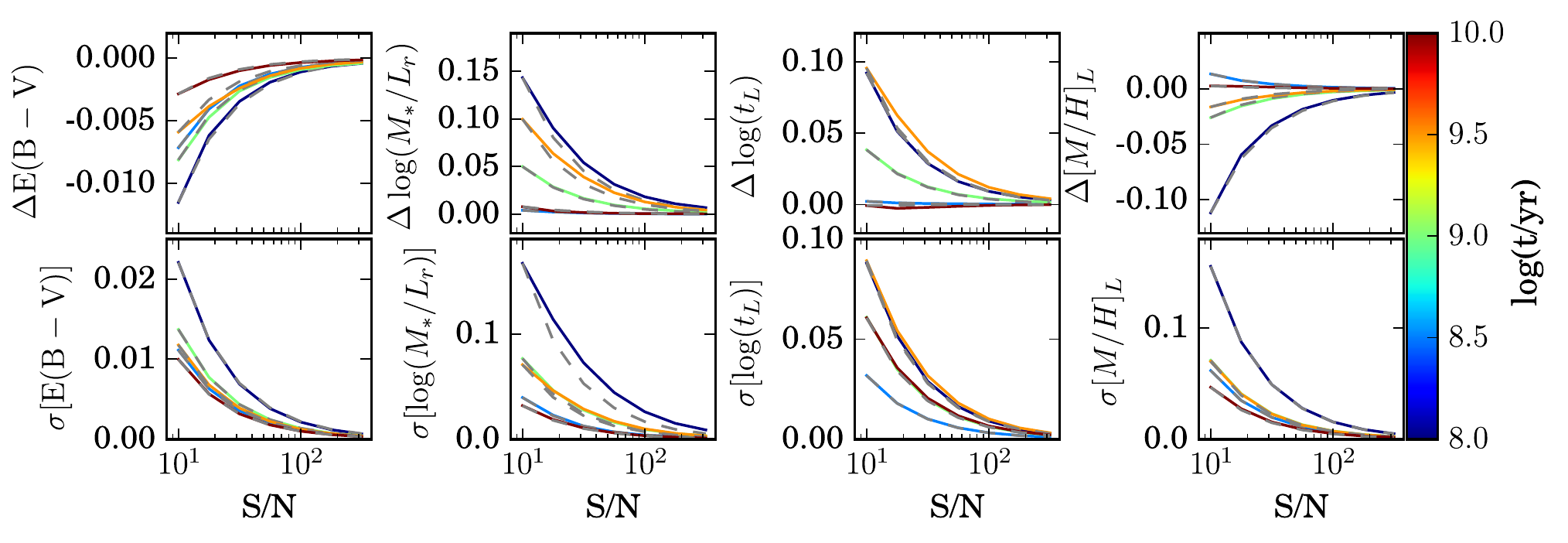}
\caption{
The parameter bias (top four panels) and scatter (bottom four panels) as a function of S/N for pPXF fitting based on 
1000 simulations of each spectrum. The colored lines correspond to different stellar ages. The grey dashed
line in each panel shows the derived relation of $P=k_P \times $1/(S/N), where $P$ corresponds to the parameter bias/scatter
at each SSP age, and $k_P$ shows the coefficient between $P$ and 1/(S/N) at S/N=10.
These parameter bias and scatter can be well described by $k_P \times$1/(S/N) in most cases, and the corresponding coefficients
are listed in Table \ref{s2n_coef}.
With these coefficients we can estimated the parameter bias and scatter based on 1/(S/N), which can be easily measured.
Here the parameter bias corresponds to the 50th percentile value of each parameter bias, the parameter scatter
is defined as: $\sigma(P)=\frac{1}{2}(\Delta P_{\rm 84th}-\Delta P_{\rm 16th})$,
where P corresponds to E(B-V), $M_*/L_r$, $\log t_L$, and [M/H]$_L$, $\Delta P_{84th}$ and $\Delta P_{16th}$ correspond to
the parameter bias value at 84th and 16th percentiles, respectively. 
}
\label{ppxf_par_vs_snr}
\end{figure*}

With the selected full-spectrum fitting codes and SSP library, we first test the impact of 
S/N and error spectra variation on the fitting results. These two tests provide guidance 
when analyzing IFU data, on how to select S/N thresholds for Voronoi binning, 
and how much biases and scatters are expected given different S/N's and error spectral shapes.
These tests will be done for a single metallicity. 
Once we are clear about these two effects, we adopt a single set of S/N and spectral error type to test 
the systematic bias and scatter of the measured stellar population parameters for the whole range of metallicities.

\subsection{Uncertainties and systematics introduced by measurement noise}
\begin{figure*}
\centering
\includegraphics[angle=0.0,scale=0.95,origin=lb]{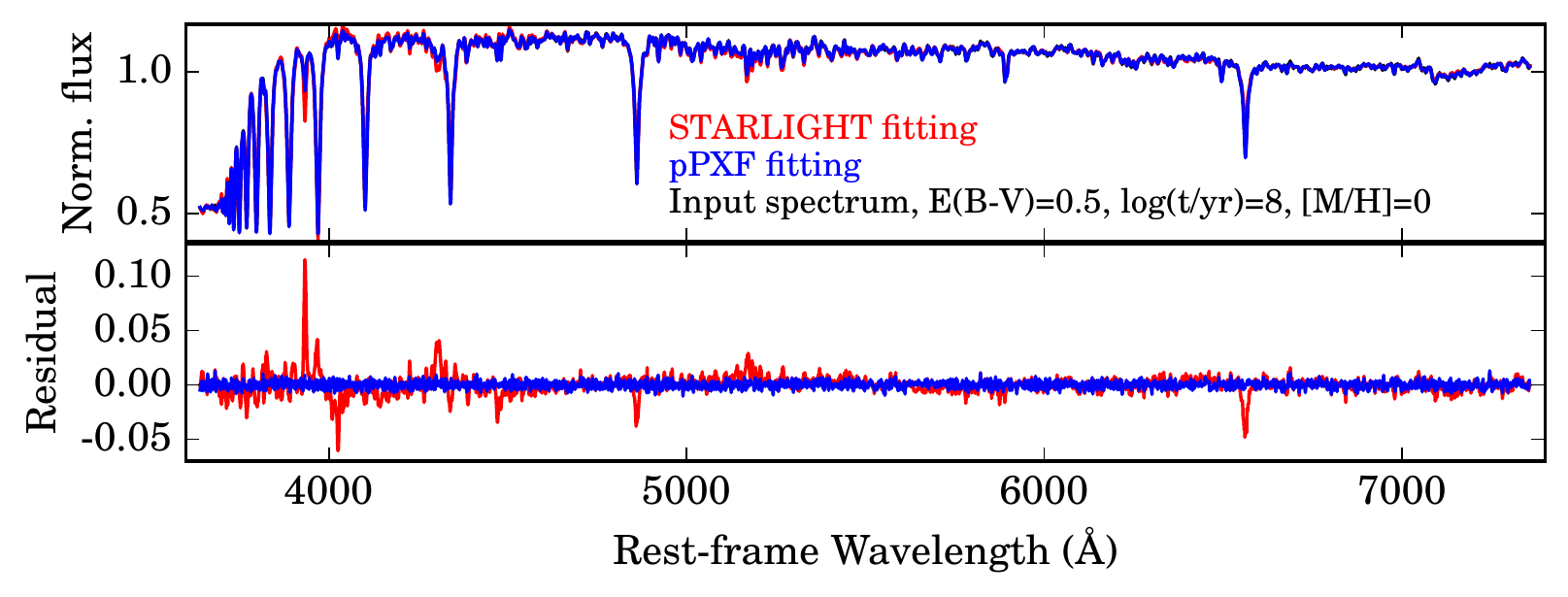}\\
\includegraphics[angle=0.0,scale=0.75,origin=lb]{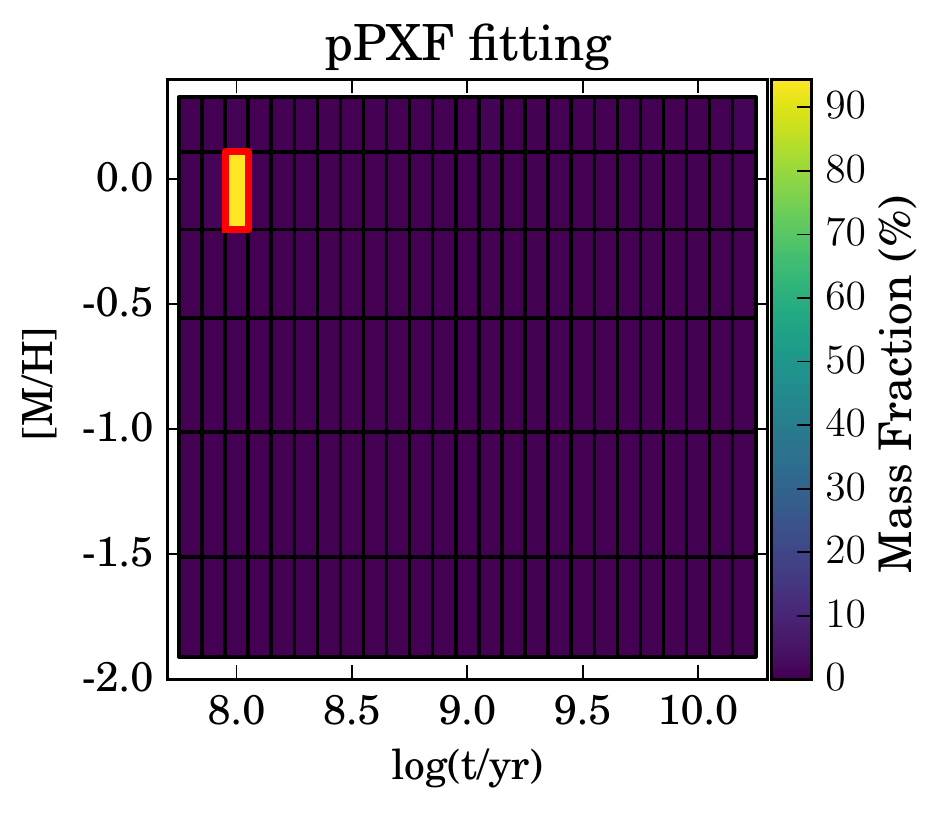}
\includegraphics[angle=0.0,scale=0.75,origin=lb]{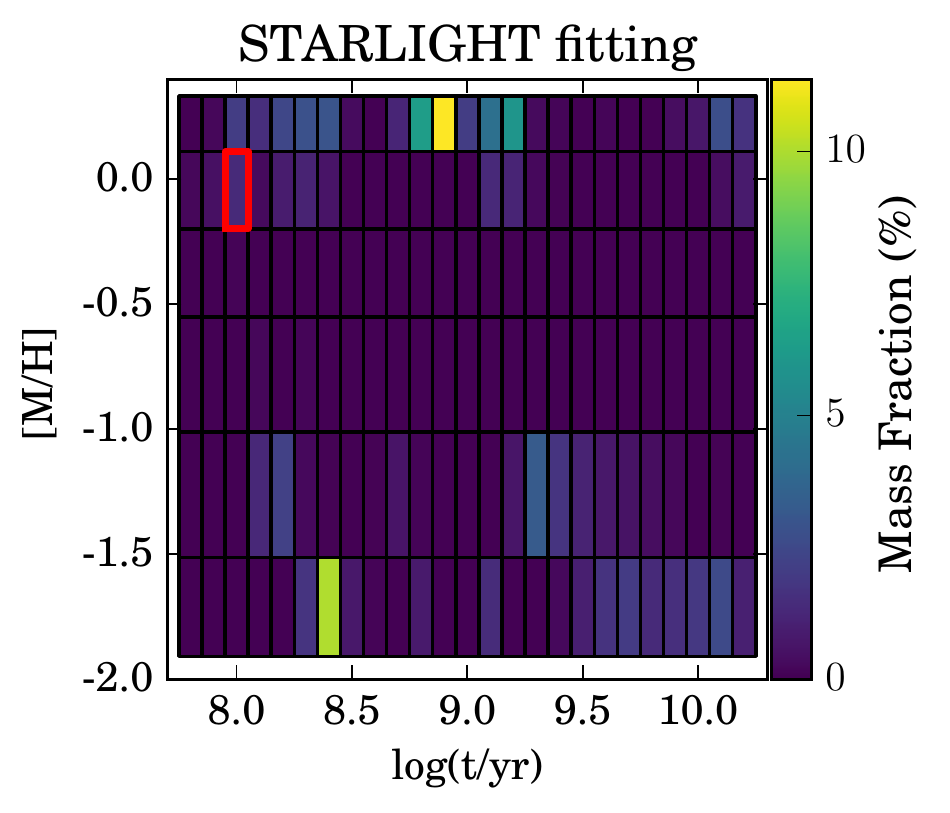}
\caption{
An example of the full-spectrum fitting to an SSP with solar metallicity, $10^8$ yr age, input E(B-V)=0.5 and S/N=316. 
The fitting results and residuals by pPXF and STARLIGHT codes are shown in the top panel. The black color shows the input spectrum, the red 
color shows the fitted model spectrum and residual by STARLIGHT, while the blue color represents the model spectrum and residual by pPXF.
The bottom left panel shows the averaged SFH map obtained from 50 Monte-Carlo simulations by pPXF fitting, 
while the bottom right panel shows the same SFH map by STARLIGHT fitting. Each SSP is enclosed by a black box,
and the color in each black box represents the mass fraction of each SSP.
The solid red box shown in each SFH map labels the position of the input SSP.
}
\label{spec_sfh}
\end{figure*}

To derive the parameter bias and scatter under different spectral S/N's, we select five SSPs with different
ages ($\log (t/{\rm yr})=$8.0, 8.5, 9.0, 9.5, 10.0) but the same metallicity ([M/H]=0) for simulation. To generate the mock data,
we assume a flat error spectrum in both linear (for STARLIGHT) and log (for pPXF) binned wavelength, 
and seven S/N's equally-sampled in logarithmically-space between 1.0 to 2.5 
(S/N=10, 18, 32, 56, 100, 178, 316):
\begin{equation}
	S/N = \frac{1}{N_p} \sum_i S_{\lambda i}/N_{\lambda i}, \lambda_i=[5490, 5510]\AA.
\end{equation}
where $N_p$ is the number of wavelength pixels included in [5490, 5510]\AA.

In our analysis, we mainly focus on the bias and scatter of E(B-V), $M_*/L_r$, age ($\log t$) and metallcity ([M/H]).
For each mock spectrum, we perform 50 Monte-Carlo simulations by assuming that the flux uncertainties
at each wavelength point follow a Gaussian distribution. After the full-spectrum fitting of each simulated spectrum, 
we can measure the following population parameters: $r$-band stellar mass-to-light ratio ($M_*/L_r$), luminosity-weighted age ($t_L$), and
metallicity ($[M/H]_L$) as follows:
\begin{equation}
M_*/L_r=\frac{\Sigma f_{M,i}}{\Sigma f_{M,i}/(M_*/L_r)_i}
\end{equation}
\begin{equation}
\log(t_L)=\Sigma f_{L,i}\times\log(t_i/yr),
\end{equation}
\begin{equation}
[M/H]_L=\Sigma f_{L,i}\times[M/H]_i. 
\end{equation}
where $(M_*/L_r)_i$, $\log(t_i/yr)$ and $[M/H]_i$ correspond to the $r$-band mass to light ratio, age, and metallicity 
of the $i$-th SSP, while $f_{L,i}$ and $f_{M,i}$ are the fitted luminosity and mass fractions, respectively. 
We can then calculate the parameter biases for each spectrum based on 50 simulations:
\begin{equation}
\Delta E(B-V) =  \frac{1}{N} \sum_i E(B-V)_i-E(B-V)_{\rm input}, 
\end{equation}
\begin{equation}
\Delta \log (M_*/L_r) =  \frac{1}{N} \sum_i \log(M_*/L_r)_i-\log(M_*/L_r)_{\rm input}, 
\end{equation}
\begin{equation}
\Delta \log t_L= \frac{1}{N} \sum_i \log t_{L,i}-\log t_{\rm input}, 
\end{equation}
\begin{equation}
\Delta[M/H]_L= \frac{1}{N} \sum_i [M/H]_{L,i}-[M/H]_{\rm input}. 
\end{equation}

Considering that the measured parameter scatter can be non-Gaussian, we hence select the 16th and 84th 
percentiles of the measured parameter distribution from 50 simulations as $1\sigma$ error bars.

Figure \ref{st_snr} shows the fitting bias of measured E(B-V), $M_*/L_r$, age and metallicity obtained from STARLIGHT fitting,
population parameters are recovered well when the input E(B-V) is less than 0.2, which is consistent with
the test results shown in Figure 4 of \cite{cid2005}, whose simulation covers the input $A_V\leq 0.5$.
At E(B-V)$< 0.2$, the parameter biases and scatters decrease with increasing S/N when S/N$<60$,
especially for the cases of $\Delta\log(M_*/L_r)$ (second row) and $\Delta\log(t_L)$ (third row) at
$\log(t/{\rm yr})=$ 8.0 (first column), 9.0 (third column), and 9.5 (fourth column). 
However, when fitting those mock spectra with input E(B-V)$\geq$0.2, the bias of the four
parameters does not systematically decrease with increasing S/N and E(B-V), which is clearly shown in the $\log(t/{\rm yr})=8$ case:
the parameter biases are roughly the same at S/N$<$60, then increase when S/N$>$60 and input E(B-V)$>0.2$.
The same trends are found for the $\log(t/{\rm yr})$=8.5, 9.0 and 9.5 cases when S/N$>$60. 
For the old stellar population ($\log(t/{\rm yr}) = 10$), STARLIGHT can recover the population parameters 
well. There is almost no bias and the results are consistent with different input E(B-V), 
and their scatters decrease with increasing S/N.

Figure \ref{ppxf_snr} shows the biases and scatters of the four parameters from the pPXF fitting,
which decrease uniformly with S/N and show no obvious variation for all input E(B-V) cases.
For those spectra with low S/N, E(B-V) are maximumly underestimated by a mean level of $\sim$0.01 magnitude.
The positive biases in $M_*/L_r$ and $\log t_L$ are due to the fact that at low S/N one can 
easily hide a significant amount of mass in old populations, 
and the resulting redder spectrum shape can be made bluer by decreasing the reddening. This old population will increase 
the $M_*/L_r$ without contributing much to the light. The negative bias in [M/H] can be understood as due to the 
age-metallicity degeneracy (Worthey 1994), which states that ``the spectrophotometric properties of an unresolved stellar population 
can not be distinguished from those of another population three times older and with half the metal content". 

Interestingly, although the population errors tend to be larger for younger populations, the increase is not monotonic. 
In fact the errors at $\log (t/\rm yr)$=8.5 are actually smaller than those at $\log (t/\rm yr)$=9. This can be understood as the 
competition between two effects: (i) young populations have strong Balmer lines to better constrain the model fit but 
(ii) young populations make it more difficult to constrain the presence of a possible old one. A sweet spot 
seems to be achieved between $\log (t/\rm yr)=8-9$.

Since the parameter bias and scatter can converge uniformly and are independent of the input E(B-V), we then perform 1000 simulations 
for each spectrum in each input E(B-V) case,
and calculate the global 16th and 84th percentiles of the 6000 (1000$\times$6 input E(B-V)) measured parameter biases.
The grey shaded region of each panel in Figure \ref{ppxf_snr} corresponds to the global 16th and 84th percentiles at each S/N.
To get a clearer idea of how these parameter biases and scatters vary with S/N, we plot the global median bias and scatter versus S/N
in Figure \ref{ppxf_par_vs_snr}. Spectra with different ages (shown in different colors) tend to show levels of biases (top four panels) 
and scatters (bottom four panels). The bias and scatter are the largest for the youngest SSP and smallest for the oldest SSP. 
At the same time, we find that the trends of these parameter biases and scatters are described well by the expected inverse 
dependency on S/N: $P=k_P \times$1/(S/N),
where $P$ corresponds to the bias/scatter at different SSP ages in each panel. The grey line in each panel shows the derived
parameter bias/scatter for each age based on this scaling at S/N=10.
We list all the $k_P$ coefficients in Table \ref{s2n_coef} so that one can estimate the pPXF fitting bias/scatter for different
ages.

\begin{table}
\caption{$k_P$ at different ssp ages in the correlation: $P = k_P\times$1/(S/N)}
\centering
\begin{tabular}{|c|c|c|c|c|}
\hline
$\log(t_{\rm SSP}/\rm yr)$  & $k_{\Delta \rm E(B-V)}$ & $k_{\Delta \log(M_*/L_r)}$ & $k_{\Delta \log t_L}$ & $k_{\Delta {\rm [M/H]}_L}$ \\
\hline
8.0 & -0.12 & 1.43 & 0.92 & -1.12 \\ 
\hline
8.5 & -0.07 & 0.04 & 0.02 & 0.13 \\ 
\hline
9.0 & -0.08 & 0.50 & 0.38 & -0.26 \\ 
\hline
9.5 & -0.06 & 1.00 & 0.95 & -0.16 \\ 
\hline
10.0 & -0.03 & 0.08 & -0.01 & 0.02 \\ 
\hline
    & $k_{\sigma \rm [E(B-V)]}$ & $k_{\sigma [\log(M_*/L_r)]}$ & $k_{\sigma (\log t_L)}$ & $k_{\sigma {\rm [M/H]}_L}$ \\
\hline
8.0 & 0.22 & 1.67 & 0.88 & 1.55 \\ 
\hline
8.5 & 0.11 & 0.40 & 0.32 & 0.62 \\ 
\hline
9.0 & 0.14 & 0.77 & 0.61 & 0.71 \\ 
\hline
9.5 & 0.12 & 0.71 & 0.89 & 0.70 \\ 
\hline
10.0 & 0.10 & 0.32 & 0.61 & 0.47 \\ 
\hline
\end{tabular}
\label{s2n_coef}
\end{table}
\begin{figure*}
\centering
\includegraphics[angle=0.0,scale=0.95,origin=lb]{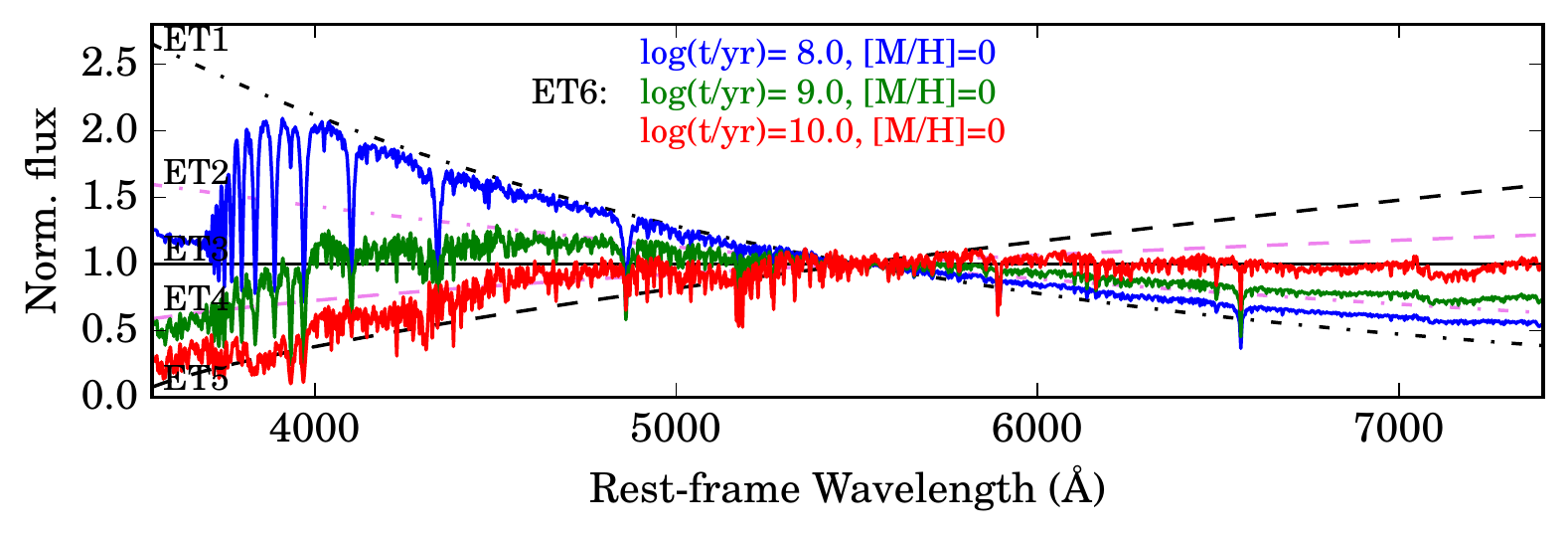}
\caption{
Six selected spectral error types for the flux uncertainty to test full-spectrum fitting algorithms. 
The five smooth curves are error type 1 (ET1) through error type 5 (ET5).
ET6 is defined to have constant S/N at all wavelengths. Thus, ET6 varies with input spectrum.
The three colored error spectra show ET6 error spectra for three different SSPs
with $\log(t/{\rm yr})=$ 8.0, 9.0 and 10.0, and solar metallicity.
}
\label{err_setup}
\end{figure*}
\begin{figure*}
\centering
\includegraphics[angle=0.0,scale=0.8,origin=lb]{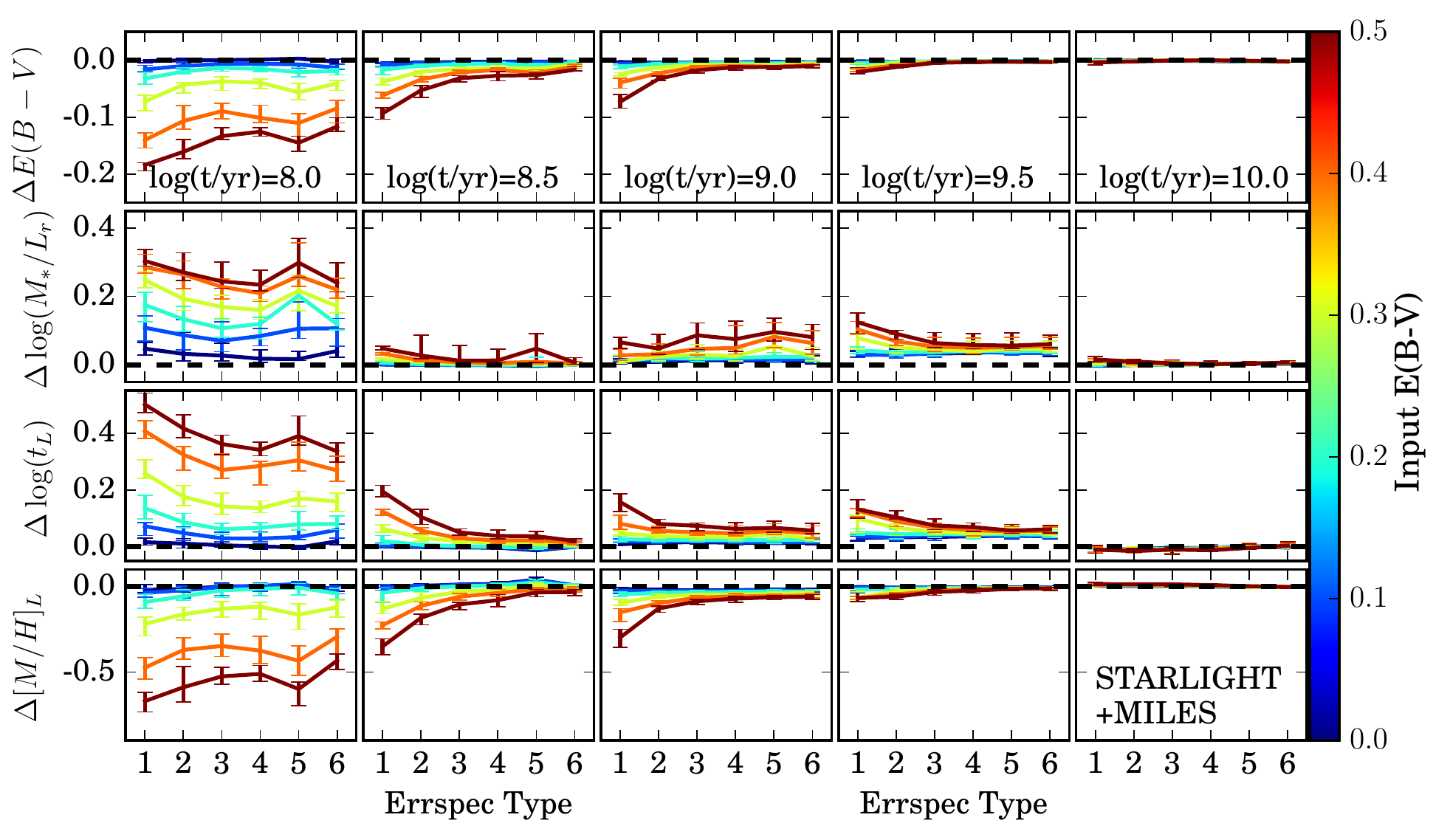}
\caption{
STARLIGHT fitting results at S/N=60 obtained by using different error spectral types as shown in Figure \ref{err_setup}. 
Errspec Type 1 to 6 represent ET1 to ET6, respectively. The input E(B-V) and SSP age setup are the same as 
shown in the S/N test section (Figures \ref{st_snr} and \ref{ppxf_snr}).
}
\label{err_test_st}
\end{figure*}
\begin{figure*}
\centering
\includegraphics[angle=0.0,scale=0.8,origin=lb]{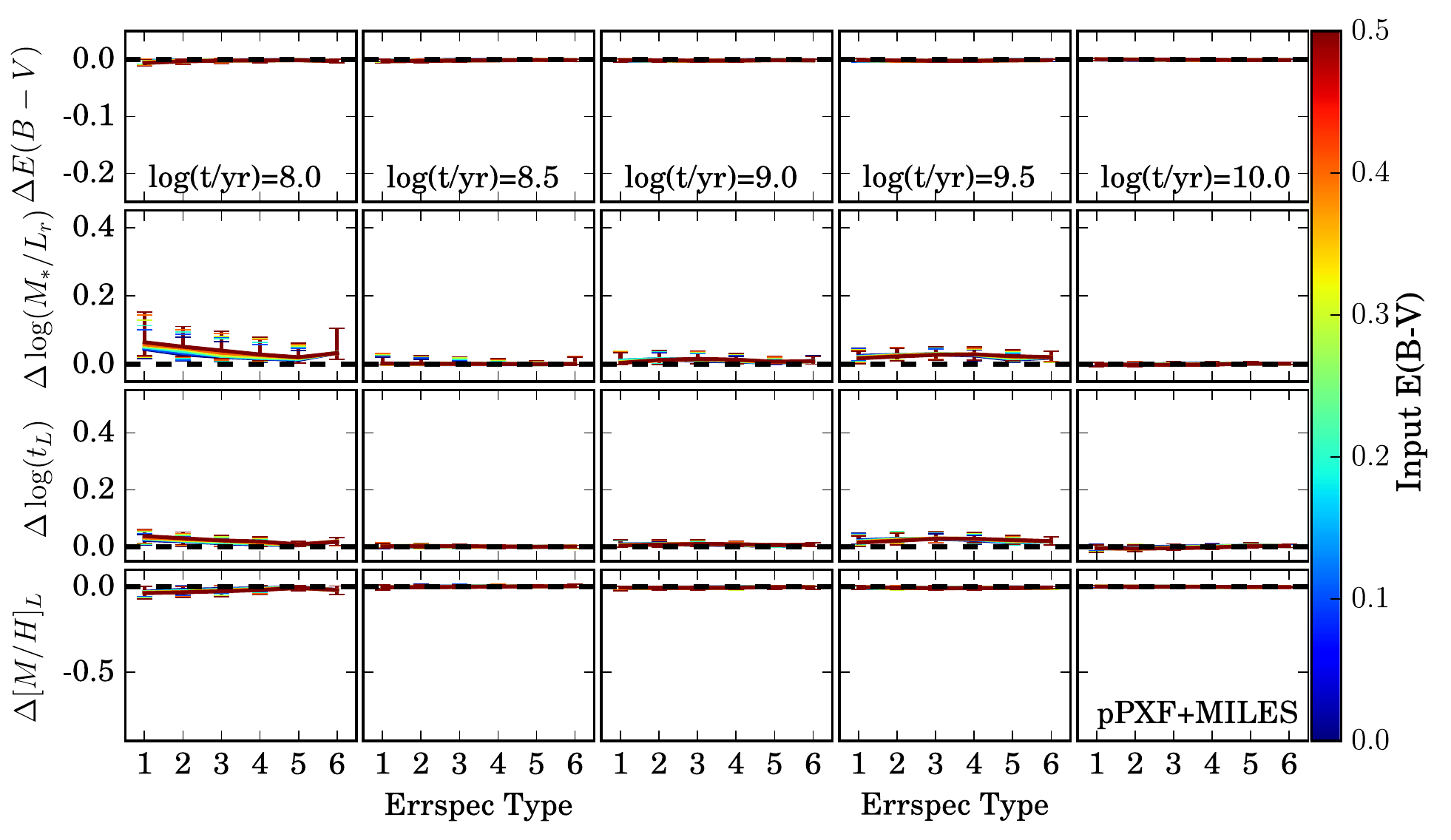}
\caption{
The variation of pPXF fitting results at S/N=60 with different error spectrum types, the labels and colors are 
the same as shown in Figure \ref{err_test_st}.
}
\label{err_test_ppxf}
\end{figure*}
To have a better understanding of what is causing the fitting bias, we show an example of spectral fitting details in Figure \ref{spec_sfh}.
The mock spectrum is generated based on an SSP with $\log(t/{\rm yr})=8$, solar metallicity, and input E(B-V)=0.5. 
At high S/N (=316) shown in Figures \ref{st_snr} and \ref{ppxf_snr}, 
STARLIGHT still has parameter biases, while pPXF essentially only recovers as output the single input spectrum, 
as expected at this extreme S/N,
with a bias and scatter less than $\sim$0.01 dex (Figure \ref{ppxf_par_vs_snr}).
From the residuals we can see that pPXF yields a better fit to both the absorption-line and continuum features,
but the fitting by STARLIGHT shows significant residuals for many absorption lines. 

The fitting performance can be further clarified by the SFH distribution shown in the bottom two panels of Figure \ref{spec_sfh}.
The pPXF solution essentially consists of a single component, the input SSP with 95\% of the weight, plus some minimal numerical noise.
For STARLIGHT, there are significant contributions from $\sim 1$ Gyr SSPs, which explains the large biases 
shown in Figure \ref{st_snr}. 

Based on the above analysis, for pPXF fitting, the parameter biases decrease with increasing S/N.
For STARLIGHT, the measured parameters converge to the input values only for cases with E(B-V)$<$0.2,
and show larger biases with increasing S/N for E(B-V)$>$0.2. In the E(B-V)=0.5 and $\log(t/\rm yr)=8$ or 8.5 cases, 
the increasing of parameter bias as a function of S/N starts from $\sim$S/N=60. Therefore, applying S/N=60 for further
analyses tends to be a reasonable choice, those parameter biases at different S/N's can be estimated easily for both
pPXF and STARLIGHT.
Next, we test the dependence of the fitting results on the shape of the error spectra with spectral S/N at $\lambda = [5490, 5510]$\AA..

\subsection{Impact of error spectrum shapes}

All observed spectra come with uncertainties in the flux. Full spectrum fitting codes take these 
uncertainties into account during fitting. This potentially leads to different spectrum regions being 
weighted differently in deriving the results. Could different wavelength dependence in these weights 
lead to different biases in the fitted parameters? Here we check the dependence of spectral fitting results
on the shape of the error spectrum. 

Six kinds of error spectra are designed based on the spectral types of SSPs and MaNGA error spectral types. 
In Figure \ref{err_setup}, ET1 and ET5 are consistent with the continuum slope of a young ($10^8 yr$)
and an old ($10^{10} yr$) SSP with solar metallicity, respectively. 
ET3 represents a flat error spectrum which has already been used in the S/N test above. 
ET2 is half way between ET1 and ET3, and is similar to the error spectra of typical MaNGA data. 
ET4 is half way between ET3 and ET5. ET6 is the case of a constant S/N per wavelength pixel, 
three spectral examples with stellar age $\log(t/{\rm yr})=$8.0, 9.0 and 10.0 and solar metallicity
are shown in blue, green and red colors.

Figure \ref{err_test_st} shows the test results for STARLIGHT with these six types of error spectra.
When the error type varies from ET1 to ET5, the measured parameter biases become smaller, especially 
for those spectra with high dust extinction. The ET1 error spectrum, which has the largest flux uncertainty in
the blue band than the other five types, causes the largest parameter bias. The ET5 error spectrum, which has
the smallest flux uncertainty in the blue band, induces the smallest parameter bias. These biases caused by error
spectral shapes become smaller with increasing stellar age, and disappear at $t=10^{10}$yr. 

As shown in Figure \ref{err_setup}, many absorption features are concentrated in the blue band (e.g. $\lambda < 5000\AA$), 
especially for those young SSPs (e.g. the spectrum for $t=10^8 \rm yr$).
Therefore, if there are higher flux uncertainties included in the blue band, the larger parameter biases and scatters are induced
for STARLIGHT. The fitting biases and scatters decrease with increasing stellar ages, which correspond to 
increasing absorption features included in the red band.

The performance of spectral fitting with the ET6 error spectrum also verifies the above interpretation. For the $t=10^8$yr case,
ET1 and ET6 differs mainly at $\lambda<4000\AA$. The flux uncertainty of ET6
at $\lambda<4000$\AA~ is smaller than that of ET1, the parameter biases resulted from fitting with ET6 are smaller than those with ET1.

Compared to the STARLIGHT fitting, the pPXF fitting results show very weak or no dependence on the
error spectrum types (Figure \ref{err_test_ppxf}) --- the biases and scatters are all similar in the six cases.

After checking the effects of error spectral shapes in STARLIGHT and pPXF, we select the flat 
error spectrum (ET3) for studying the performance of the two codes further in the age-metallicity parameter space
and two components based SFH tests.
Although ET2 is closer to observations (at least for the MaNGA survey), the error spectral shapes still have large
variations for different observational instruments. Therefore, applying the flat error spectrum for fitting tests would be
a reasonable choice, after which we can tell whether the fitting with an observed error spectrum will
have a larger or smaller bias compared to our fiducial case.

\subsection{Code tests with single-SSP based mock spectra}
With the understanding of S/N and error spectral type effects, we can now assume a suitable S/N (=60)
and flat error spectrum (ET3) to check the algorithm bias and scatter in the age and metallicity
parameter space of the Vazdekis/MILES SSP library. 
Mock spectra are generated based on single SSPs with the initial setup as described in Section 2.3. 
By analyzing the fitting with mock spectra generated by single SSPs, we can have a thorough interpretation on 1) which kinds of spectra are easily 
biased, 2) when the fitting results show unavoidable biases, and 3) how large these biases and scatters are.

\begin{figure*}
\centering
\includegraphics[angle=0.0,scale=0.8,origin=lb]{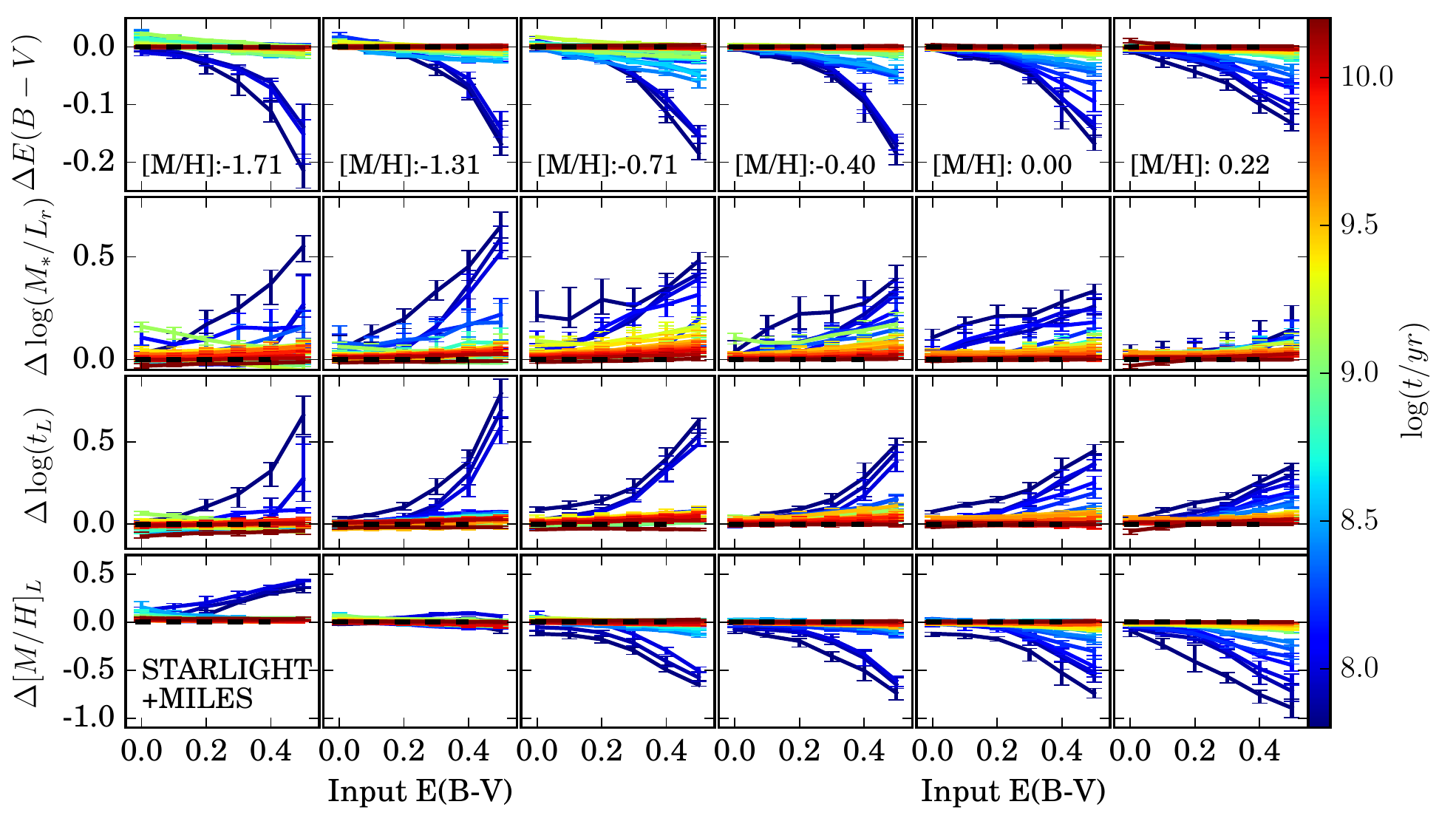}
\caption{
STARLIGHT fitting of mock spectra generated by a single SSP with different ages and metallicities  
at S/N=60.
From left to right, we show the parameter biases
at different [M/H] (-1.71, -1.31, 0.71, -0.4, 0.0, 0.22) bins. Blue to red colors represent
the stellar age ranging from 0.063 to 15 Gyr. The biases in the four parameters ($\Delta E(B-V)$, 
$\Delta \log(M_*/L_r)$, $\Delta \log t_L$, and $[M/L]_L$) are shown from top to bottom.
The zero-biased line of each parameter in each panel are labeled as the horizontal dashed line.
}
\label{st_single_ssp}
\end{figure*}
\begin{figure*}
\centering
\includegraphics[angle=0.0,scale=0.8,origin=lb]{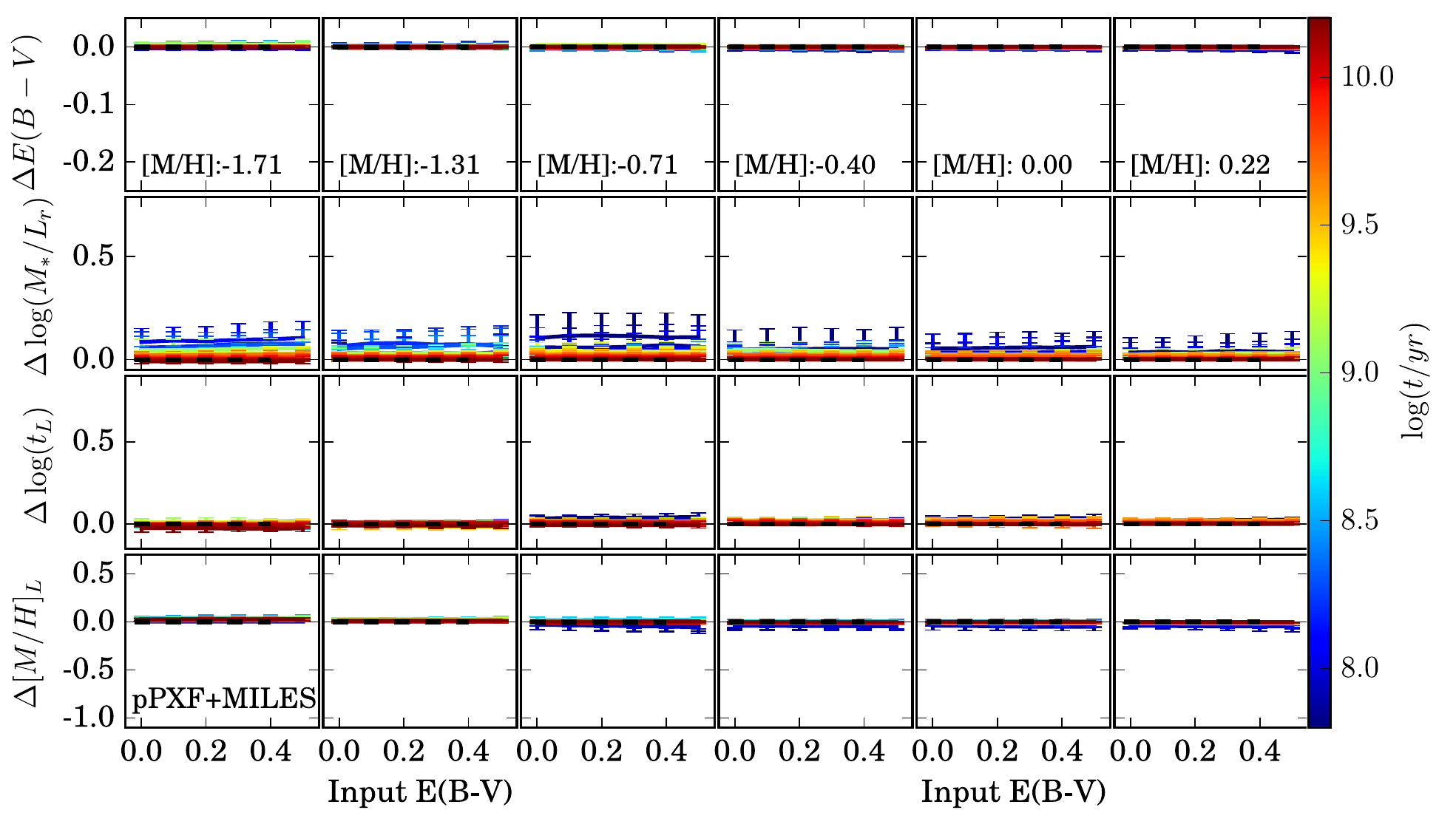}
\caption{
The pPXF fitting of mock spectra generated by a single SSP as a function of the stellar age and 
metallicity at S/N=60. Lines in different panels are the same as shown in Figure \ref{st_single_ssp}.
}
\label{ppxf_single_ssp}
\end{figure*}

Figure \ref{st_single_ssp} shows the STARLIGHT fitting results in different metallicity bins with
stellar ages labelled with rainbow colors. According to this plot we summarize three typical behaviors of the STARLIGHT fitting:
1) For those SSPs with $t<10^{9}$yr,
the measured E(B-V) biases increase with increasing dust extinction for all metallicities. 
2) Stellar population parameter biases increase with younger stellar ages.
3) For those SSPs with $t>10^9$yr, the fitted population parameters show consistent results with the input values.

When fitting young SSPs with significant extinction, STARLIGHT invokes older SSPs to fit the red continuum shape, 
while pPXF has no problem finding the correct extinction. 

As shown in the top six panels of Figure \ref{ppxf_single_ssp}, the dust extinction measurements from pPXF are 
insensitive to the input E(B-V), no matter what stellar metallicity is adopted.
However, for each metallicity, spectra with younger ages always result in larger $M_*/L_r$ biases, which is
caused by the contamination of old populations that contributes little to the light but significantly to the mass.
These artificial old SSPs have little effects on the light-weighted ages and metallicities, but can introduce parameter scatters as shown 
in the third and fourth rows of Figure \ref{ppxf_single_ssp}.

\begin{figure*}
\centering
\includegraphics[angle=0.0,scale=0.7,origin=lb]{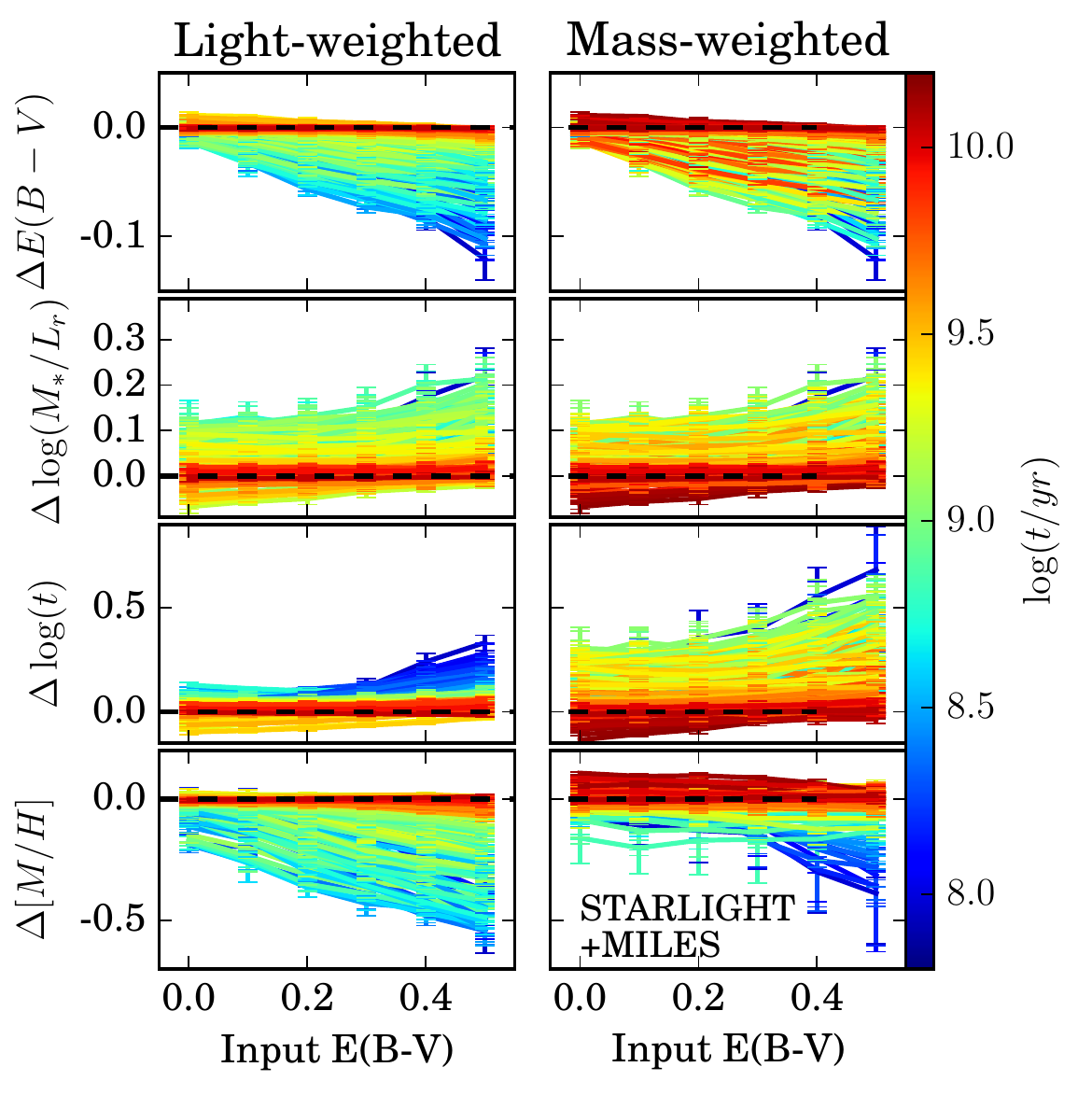}
\includegraphics[angle=0.0,scale=0.7,origin=lb]{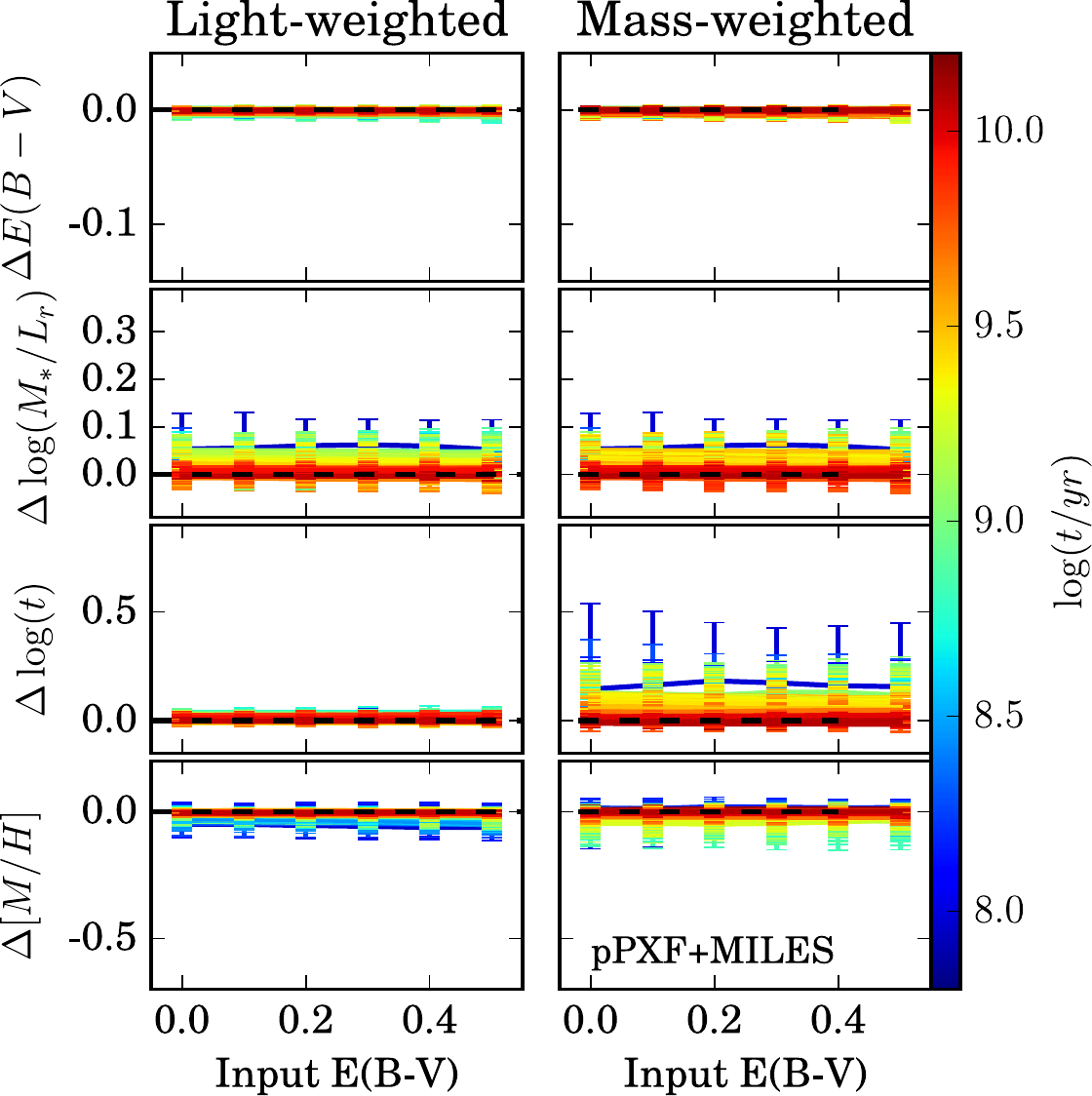}
\caption{
Full-spectrum fitting results of two-component SSPs co-added mock spectra at solar metallicity with S/N=60. The first and 
second columns show the light-weighted and mass-weighted results by STARLIGHT, while
the third and fourth columns show the corresponding results by pPXF. Colors in the first and third columns 
represent the light-weighted ages, while those colors in the second and fourth columns correspond to 
mass-weighted ages. For E(B-V) and M*/Lr (the first and second rows), the curves are identical between 
light-weighted and mass-weighted cases, although they are colored differently by the input ages.
The zero-biased line of each parameter is labeled with a horizontal dashed line.
}
\label{2comb_fit}
\end{figure*}

\subsection{Linear combination of two different SSPs}

Considering that the SFH of a galaxy is usually not dominated by a single SSP, 
we test the performance of the pPXF and STARLIGHT codes by combining two different SSPs with solar metallicity. 
We take a set of 13 SSPs with solar metallicity, logarithmically spaced in age between 0.063 and 15.8 Gyr 
(i.e. we skip every other age from our full set of SSPs), and consider all 78 combinations of two SSPs without repetition.
The two selected SSPs are co-added together after normalizing fluxes to 1 at $\lambda = [5490, 5510]$\AA.
Then we perform the simulation with the same steps as the single SSP tests.

The age and metallicity of two-component SSPs co-added spectra, which can have 
light-weighted values defined in Equations (2-8) and mass-weighted values defined as follows:
\begin{equation}
\log t_M=\Sigma f_{M,i}\times\log t_i,
\end{equation}
\begin{equation}
[M/H]_M=\Sigma f_{M,i}\times[M/H]_i. 
\end{equation}
we then calculate the biases of age and metallicity respectively as follows:
\begin{equation}
\Delta\log t_M=\frac{1}{N}\sum_i \log t_{M,i}-\log t_{M,\rm input}, 
\end{equation}
\begin{equation}
\Delta[M/H]_M=\frac{1}{N}\sum_i [M/H]_{M,i}-[M/H]_{M,\rm input}. 
\end{equation}
We also select the 16th and 84th percentiles of the measured parameter distribution from 50 simulations as $1\sigma$ error bars.

Figure \ref{2comb_fit} shows the fitting results of two-component SSPs co-added mock spectra
obtained from STARLIGHT (left two columns) and pPXF (right two columns), respectively.
For STARLIGHT, the light-weighted parameter biases and scatters increase at E(B-V)$<$0.3 for those spectra with $\log(t_L/{\rm yr})<9.5$.
More spectra with middle ages ($8.5<\log(t_L/\rm yr)<9.5$) have increased parameter biases ($>1\sigma$) compared to single-SSP case 
(Figure \ref{st_single_ssp}). 
For pPXF, the light-weighted results are consistent with those in the single-SSP case (Figure \ref{ppxf_single_ssp}).

During the spectral fitting, a significant fraction of mass in an old population can be hidden if it produces little light.
Therefore, the light-weighted population parameters, which are more directly linked to the light coming from the spectra, 
are always more accurate than mass-weighted ones.
This is clearly shown in the third row of Figure \ref{2comb_fit} for both pPXF and STARLIGHT, where the 
measured light-weighted stellar ages have smaller biases and scatters than mass-weighted ones.
While for the measured [M/H], the contaminating old SSPs with significant mass fractions but small light fractions 
can have both lower and higher [M/H] (see the examples shown in Figure \ref{spec_sfh}), hence the mass-weighted [M/H]
have only larger scatters but no obvious larger biases than light-weighted ones.

From the tests based on co-added mock spectra of two-component SSPs, which mimics the observed galaxy spectra better
than a single SSP, we can see that pPXF recovers the input parameters well for all input E(B-V) cases. While for STARLIGHT,
those young spectra with $t<3$ Gyr tend to have larger biases with younger stellar ages. For those spectra older than 
3 Gyr, although biases still exist, their fitting results become closer to the true values.

\section{Discussion}

According to our analyses, the full-spectrum fitting results from STARLIGHT and pPXF show quite
different parameter dependences and bias trends. The pPXF code, which perform the fitting optimization
as a quadratic problem, can always converge to the best solution in a finite number of steps. 
The STARLIGHT code, which is based on Monte-Carlo Markov Chains and annealing loops, have significant 
dependences on many parameters, such as fitting weights of different absorption lines, clipping, number of Markov Chains and loops.

\subsection{STARLIGHT fitting improvement by changing line weights and clipping}
In STARLIGHT, one can mask emission lines by setting their weights to zero in the ``Masks.EmLines.SDSS.gm'' file, 
and can also give more weights to absorption lines (e.g. 10X or 20X weights to Ca II K and G-band) to improve the fit.
As shown in Figure \ref{st_snr}, the dust extinction correction becomes worse for larger input E(B-V). 
In the case of $\log (t/{\rm yr})=8$ with input E(B-V)=0.5 (left column), the fitted E(B-V) is under-estimated by 0.2 mag, 
which corresponds to $\Delta$E(B-V)=$-$0.2. 
If one gives 10 times more weight to the Ca II K and G-band, $\Delta$E(B-V) is reduced to -0.1 mag.

As shown in the bottom right panel of Figure \ref{spec_sfh}, there are many SSP components with very small
light fractions but relatively large mass fractions, which are mainly due to the flux uncertainty that makes 
the light scattered to larger ages. These components cannot be taken seriously due to their 
small light fractions (e.g. those black boxes with light fraction $<1\%$). 
However, the parameter biases are mainly caused by those SSP components with large 
light/mass fractions (e.g. $>10\%$). In the latest published STARLIGHT code,
$A_V<1$ is imposed when initializing the Markov Chains (Roberto Cid Fernandes, priv. comm.), 
which actually constrains the search of the minimum $\chi^2$ when the true $A_V$ is larger than 1 mag. 
Based on the current initialization setup, fitting spectra with $A_V<1$
can converge more easily than those spectra with $A_V>1$.
This prior explains why the spectra with input E(B-V)$\leq 0.2$ show better convergence results than those with input
E(B-V)$>0.2$. When fitting spectra with $A_V>1$, increasing the number of Markov Chains and annealing loops 
is a possible way to improve the fitting quality.

Therefore, we test a ``slow'' mode (as described at the 
end of the config file ``StCv04.C11.config'') setup of STARLIGHT
fitting in the Appendix, by increasing the number of chains and loops but keeping other parameters the same as 
shown in Section 2.3. 
When applying the ``slow'' mode parameter setup, the spectral fitting
results show significant improvements.

Based on the above analyses, the STARLIGHT fitting results can be improved by setting different priors, 
number of Markov Chains and annealing loops, etc. The current default setup, which is established already based on lots
of efforts, still requires large improvement especially on fitting those spectra with large dust 
extinctions, e.g. E(B-V)$>0.2$.

\subsection{Understanding our fitting results}
The parameter biases shown in Figure 2 and 3 can be interpreted as due to two effects: 
(1) For STARLIGHT with input E(B-V)>0.2, the results are mainly biased by the small starting 
value in E(B-V) in the current version of the code, which produces slow convergence of the chain 
for large E(B-V). This could be easily fixed in the code (Cid Fernandes private communication). 
(2) Apart from this issue, both pPXF and STARLIGHT have similar trends with S/N. The main bias is 
caused by the well-known outshining effect: when the spectrum light is completely dominated by a 
young population, at low-S/N it becomes possible to ``hide'' significant mass fractions from
old populations, due to their relatively large $M_*/L$.
This effect explains 1) why younger spectra have stronger parameter biases, and 2) why the biases in age and 
$M_*/L_r$ are always positive. Negative biases in [M/H]$_L$ are then induced, which can be explained by 
the age-metallicity degeneracy, to keep the fit unchanged. At low S/N, the noise 
washes away the differences between spectrally similar templates, which enlarge the measured parameter scatters.

The volume limitation in both age and metallicity grids is a possible reason for introducing 
parameter biases, especially when the input value is near the edge of the model grid.
If the age grid limitation dominates the age bias, then we can derive positive age bias at $\log (t/\rm yr)=8.0$ case,
and negative bias at $\log (t/\rm yr)=10.0$ case. However, we do not see this trend.
The parameter biases for $\log (t/\rm yr)=8.0$ (at the edge of age grid) 
and $\log (t/\rm yr)=9.5$ (in the middle of age grid) show similar trends (see the third column of Figure 4).
The selected solar metallicity is close to the edge of metallicity grid.
To check whether the results are limited by model grids, we do the same test as shown in Figures 2 and 3 for 
ages $\log (t/\rm yr)=8.0, 8.5, 9.0, 9.5, 10.0$ at [M/H]=$-0.4$, at the middle of the metallicity grid. 
The derived results are very similar to that shown 
in Figures 2 and 3, which suggest that the current biases are not caused by the limitation of metallicity grid.
Here we do not show the [M/H]=$-0.4$ related results as new figures because of their high similarity to Figures 2 and 3.

\subsection{Execution time comparisons}

When applying the pPXF (in Python) and STARLIGHT (in Fortran) codes for spectral fitting, 
their computation times vary greatly. The fitting times are affected by many parameters, 
such as the spectral S/N and different fitting setups (e.g. number of Markov Chains and annealing loops for STARLIGHT fitting).
For example, for the settings described above, for Mac Os X10.8, Python version 2.7.2 and GCC version 4.8.1,
the pPXF code can fit a spectrum with $\log(t/{\rm yr})=9$, [M/H]=0, input E(B-V)=0.2, S/N=60 and a flat-shape error spectrum, 
in $\sim 0.8$ seconds, while 
STARLIGHT takes $\sim 56$ seconds, which means pPXF is 70 times faster than STARLIGHT.

The STARLIGHT fitting with the ``slow'' mode takes more time to run.
Compared to the default setup, the ``slow'' mode fitting is 11 times slower than the default setup, and 770 times slower
than pPXF fitting.

Note that both pPXF and STARLIGHT are fitting in the limit of a large number (150 in this work) of parameters, 
namely the weights of the SSPs. 
The pPXF solves for these (linear) parameters using an efficient quadratic programming algorithm, 
while STARLIGHT solves these as general non-linear parameters. This implies that to speed up STARLIGHT 
significantly would require a major algorithmic change.

\subsection{Parameter biases at S/N=30}
For the current galaxy IFU surveys, spaxels around a galaxy's edge usually have $\rm S/N<<10$ (typically $\sim 1$),
which means that spatial rebinning is required to improve the corresponding S/N before the spectral fitting analysis.
Limited by spatial resolution and S/N of each spaxel, S/N=30 is an optional value \citep[e.g.][]{li2017} selected for Voronoi 2D binning \citep{cc2003}, 
which provides a higher S/N for spectral fitting and better spatial resolution for scientific analyses.

For STARLIGHT with the default setup, results obtained from S/N=30 (Figure \ref{st_1ssp_snr30}) have the same level of biases
and scatters as the S/N=60 case (Figure \ref{st_snr}), which means these results are strongly biased for those spectra 
with large dust extinction (e.g. E(B-V)$\geq$0.3) and young ages (e.g. $t<10^9$yr). 
For pPXF at S/N=30 (Figure \ref{ppxf_1ssp_snr30}), the dust extinction can be recovered well as the S/N=60 case (Figure \ref{ppxf_single_ssp}).
Given a spectrum with $t=10^8$yr and solar metallicity at S/N=30, which is close to the worst fitting results of pPXF,
the parameter bias is 0.06 dex in $M_*/L_r$, 0.03 dex in both age and [M/H] (see also Figure \ref{ppxf_par_vs_snr}).

\section{Conclusion}
In this paper, we have examined the performance of two full-spectrum fitting algorithms (STARLIGHT and pPXF) in deriving
basic stellar population parameters. We run the most basic input-output test in the absence of model
uncertainties and with simple SFH including only one or two SSPs. We use SSPs included in Vazdekis/MILES 
stellar population library to generate mock spectra, and also use this library to do the 
spectral fitting. We use the same extinction curve in generating and fitting spectra.
This avoids model biases due to incorrect IMF, dust reddening curve, stellar evolution model, 
and empirical stellar spectral library, 
thus giving us a chance to purely study the effect of observational errors and algorithm biases on the fitting results.
We do this to set a baseline for the minimum errors one would have with these full-spectrum fitting methods.

Even for such basic tests, as soon as we add noise and extinction, the algorithms could introduce systematic bias 
to the fitted parameters. In most cases, pPXF produces better accuracy on the derived parameters than STARLIGHT, 
and is 2$-$3 orders of magnitude faster to run. 

In particular, for young and intermediate age population with significant dust extinction, STARLIGHT yields 
significant biases in the resulting parameters. This is likely due to the slow convergence of the Markov Chain 
and annealing loops. Adopting the ``slow'' mode setup (see Appendix) with a larger number of Markov Chains and annealing loops 
reduces the bias somewhat, but still not as good as pPXF. STARLIGHT fitting results also show a clear dependence 
on the shape of the error spectrum. 

The accuracy of the derived parameters by pPXF are nearly independent of the shape of the error spectrum 
and the level of dust extinction. Unlike STARLIGHT, the accuracy of parameters improves quickly with 
increasing S/N of the spectra, as expected. The systematic bias and uncertainty of the fitted parameters 
also depend sensitively on the intrinsic age of the stellar population. Spectra of younger populations 
always result in larger bias and scatter (in logarithmic space) than older stellar populations. 

We encourage users of other full-spectrum fitting methods to also conduct such basic input-output tests 
to understand the inherent bias and scatter imposed by the observational errors and the algorithm of choice. 
These sets the floor of uncertainties one can expect. They can also be used to motivate the choice of 
S/N thresholds one wants to adopt in observations and reduction of the data.

\section*{acknowledgements}
We would like to thank to the anonymous referee for the
suggestions that helped to improve this paper.
We thank R. Cid Fernandes for his useful comments on our STARLIGHT fitting results,
and A. L. de Amorim for providing us their STARLIGHT fitting setup.
This work is supported by the National Natural Science Foundation of
China (NSFC) under grant number 11473032 (JG), 11333003 (SM), 11390372 (SM, YL),
and 11690024 (YL).
RY acknowledges support by National Science Foundation grant AST-1715898.
MC acknowledges support from a Royal Society University Research Fellowship.

\appendix

\section{STARLIGHT fitting results based on larger Markov Chains and annealing loops}
In this paper, we adopt the default setup but with normalization window changed as done in de Amorim et al. (2017): 
$\rm l\_norm$ = 5635\AA, $\rm llow\_norm$ = 5590\AA, $\rm lupp\_norm$ = 5680\AA, and $A_V$ fitting range to [-0.5, 4] 
to allow negative $A_V$ and large enough parameter space for fitting as described in Cid Fernandes et al. (2005).
There are 7 Markov Chains and 3 annealing loops included. We then increase the number of Chains and loops
to see whether the parameter biases and scatters can be reduced.

\begin{figure*}
\centering
\includegraphics[angle=0.0,scale=0.8,origin=lb]{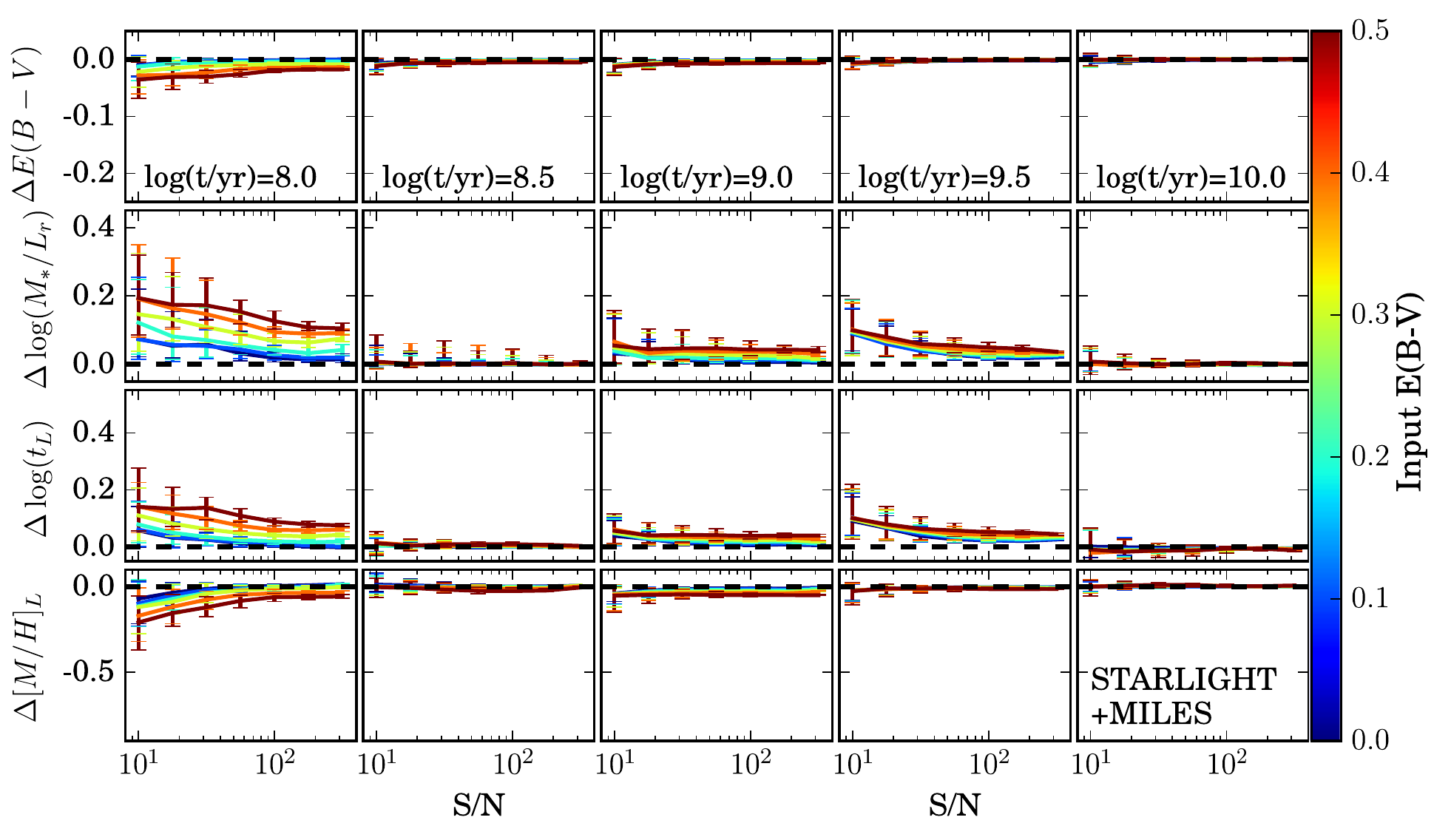}
\caption{The bias and scatter of four stellar population parameters for different spectral S/N's for STARLIGHT
based on the ``slow'' fitting mode.
The lines and colors in each panel are the same as shown in Figure \ref{st_snr}.
}
\label{apd_st_snr}
\end{figure*}

\begin{figure*}
\centering
\includegraphics[angle=0.0,scale=0.8,origin=lb]{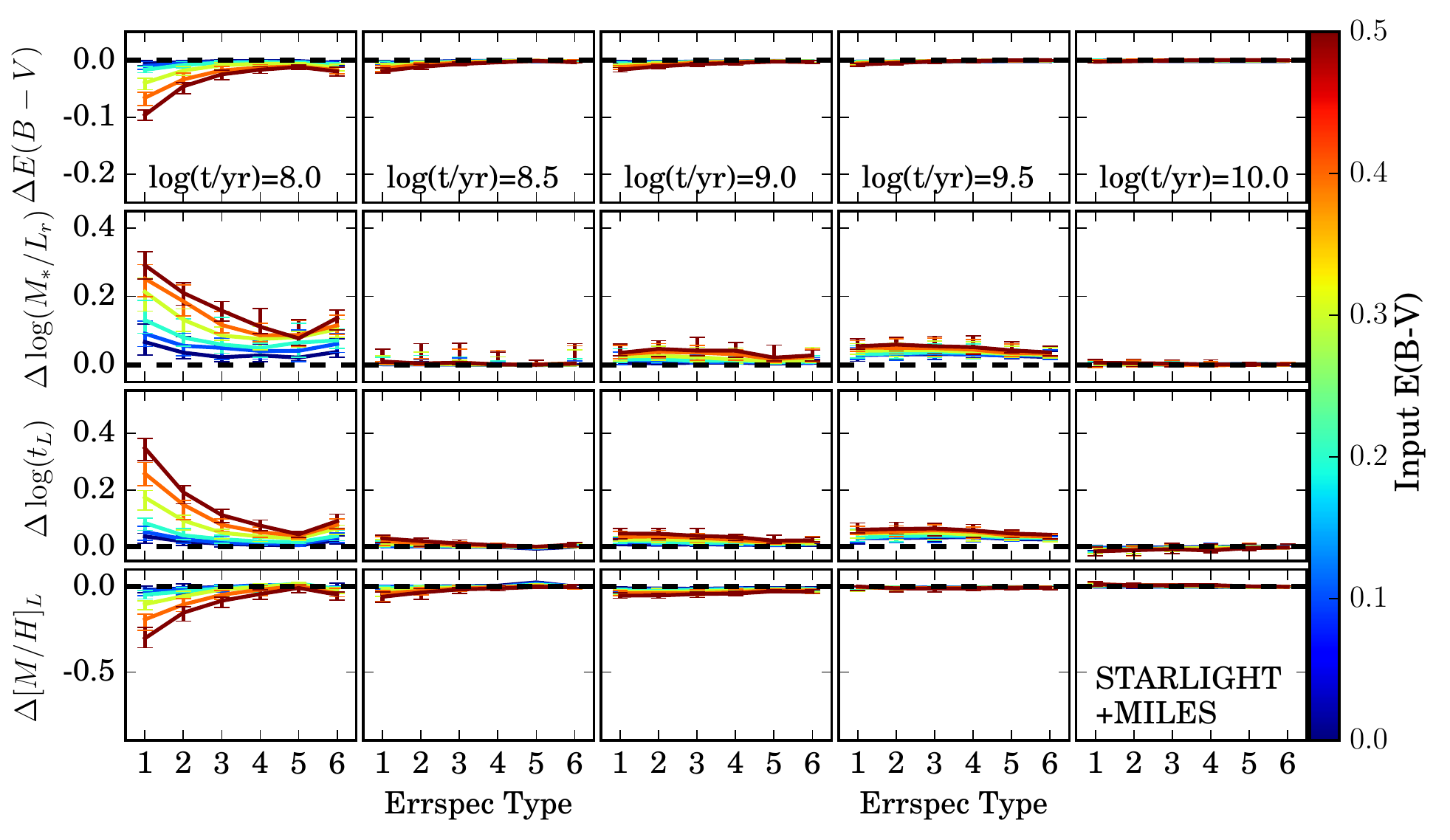}
\caption{STARLIGHT ``slow'' mode fitting results obtained with six error spectral types at S/N=60 as shown in Figure \ref{err_setup}. 
Errspec Type 1 to 6 corresponding to ET1 to ET6. The input E(B-V) and SSP age setup are the same as 
shown in the S/N test section (Figures \ref{st_snr} and \ref{ppxf_snr}).
}
\label{apd_err_test_st}
\end{figure*}

\begin{figure*}
\centering
\includegraphics[angle=0.0,scale=0.8,origin=lb]{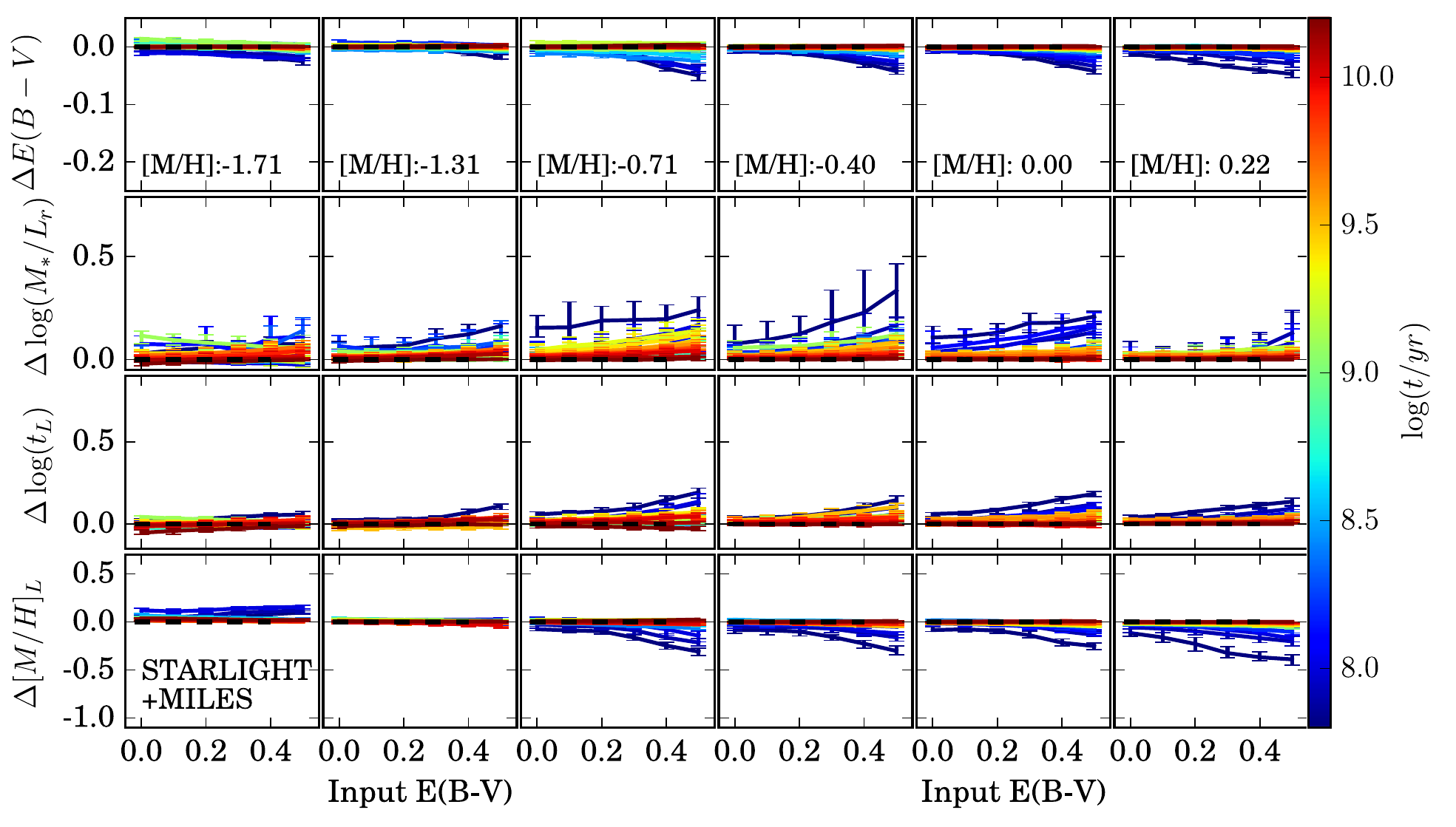}
\caption{STARLIGHT ``slow'' mode fitting of mock spectra generated by single SSP as a function of the stellar age and 
metallicity with S/N=60. Lines and colors are the same as shown in Figure \ref{st_single_ssp}.
}
\label{apd_1ssp_st}
\end{figure*}

\begin{figure}
\centering
\includegraphics[angle=0.0,scale=0.6,origin=lb]{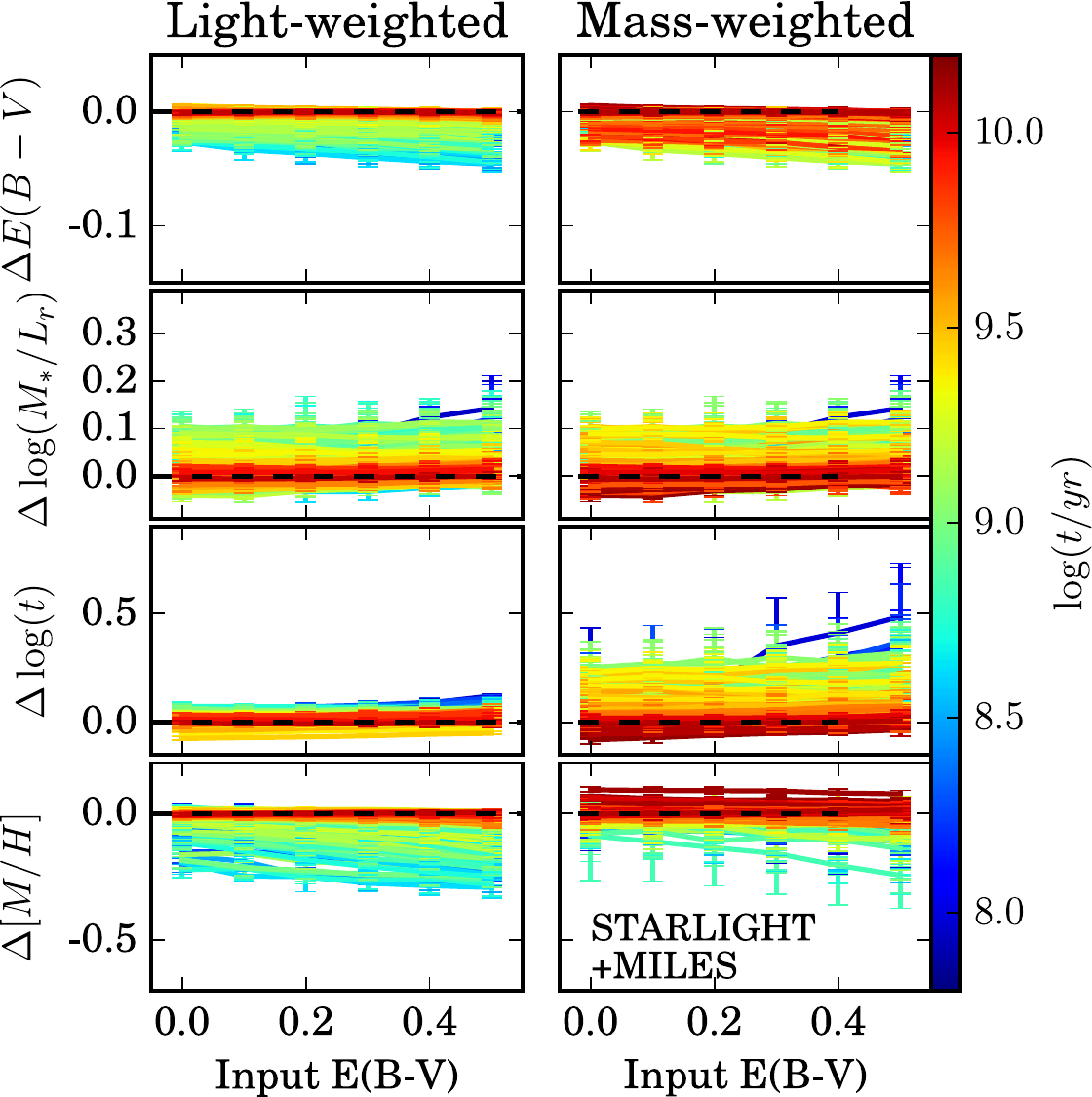}
\caption{STARLIGHT ``slow'' mode fitting results of two-component mock spectra at solar metallicity with S/N=60. 
Lines and colors are the same as shown in Figure \ref{2comb_fit}.
}
\label{apd_2ssp_st}
\end{figure}

Here we show the STARLIGHT fitting results based on 
the ``slow'' fitting mode which includes 12 Markov Chains and 10 annealing loops.
The default fitting mode is more like the ``medium'' one, which has 7 Markov Chains
and 5 annealing loops. The default setup can give similar fitting results when input E(B-V)$\leq0.2$,
which is also tested by \cite{cid2005}. While for spectra with higher dust extinction (E(B-V)$>0.2$),
the default setup will under-estimate E(B-V) much more (at least two times) than the ``slow'' mode setup.
With increased spectral S/N, the parameter biases (from the ``slow'' mode fitting) decrease for all the input
E(B-V) cases (Figure \ref{apd_st_snr}), but this decrease is less significant than that from pPXF (Figure \ref{ppxf_snr}).
The effects of error spectral types show clearer
trends (Figure \ref{apd_err_test_st}), which has already been
described in Section 3.2. When applying the ``slow'' mode parameter setup, the spectral fitting
results show significant improvements and are closer to the input for both single SSP
(Figure \ref{apd_1ssp_st}) and two-component SSP (Figure \ref{apd_2ssp_st}) tests. 

Since the STARLIGHT fitting depends on the length of chains and loops,
convergence may not be reached in the default setup.
Unlike STARLIGHT, which solves for the weights as non-linear parameters, pPXF performs the optimization
as a quadratic problem, and can quickly converge to the best solution in a small number of steps.


\section{The STARLIGHT and pPXF fitting at S/N=30}
\begin{figure*}
\centering
\includegraphics[angle=0.0,scale=0.8,origin=lb]{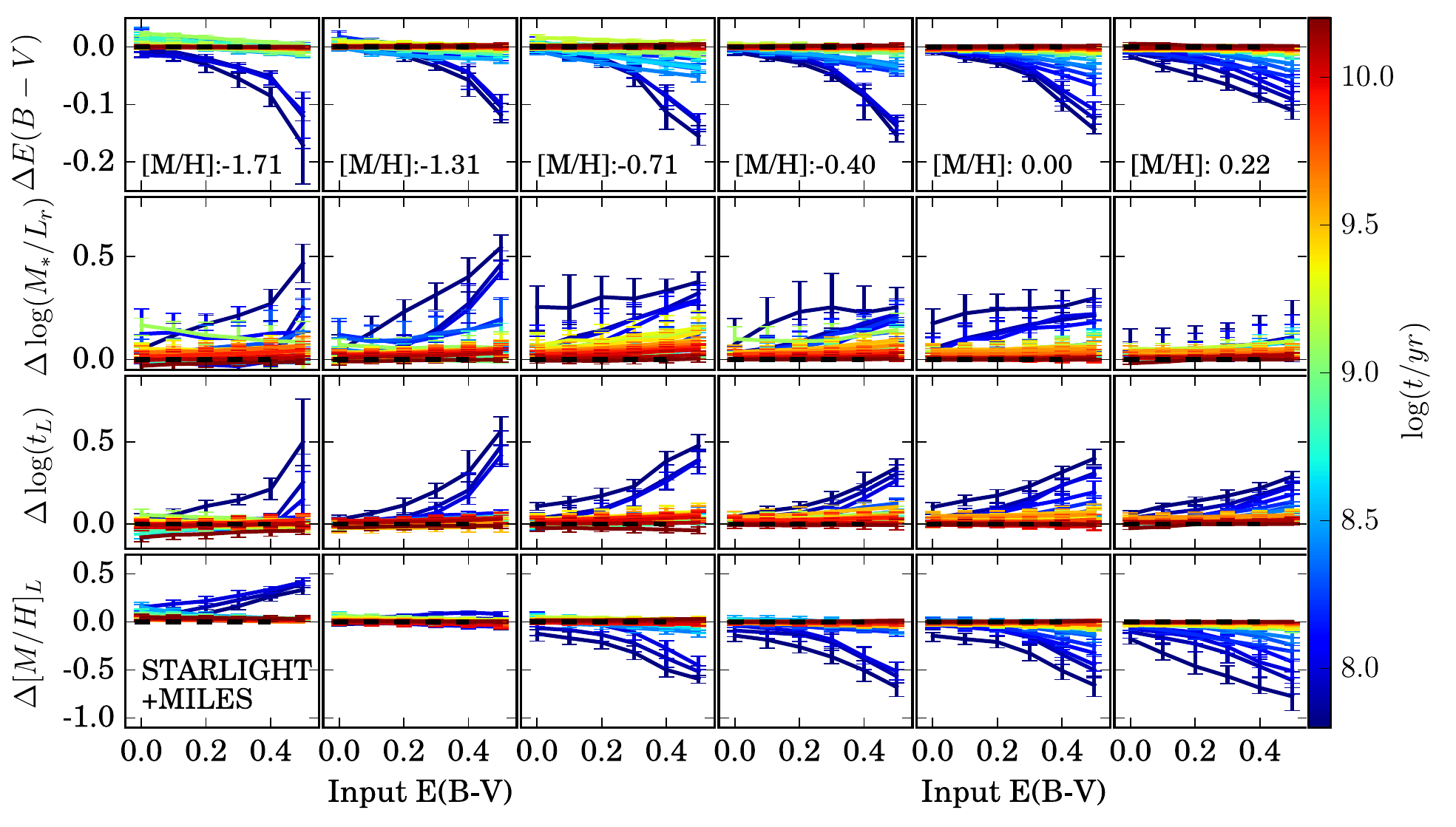}
\caption{
The STARLIGHT fitting of mock spectra generated by single SSP varies in the stellar age and 
metallicity parameter spaces with S/N=30. Lines in different panels are the same as shown in Figure \ref{st_single_ssp}.
}
\label{st_1ssp_snr30}
\end{figure*}

The STARLIGHT fitting results with the default setup at S/N=30 (Figure \ref{st_1ssp_snr30}) are similar to those at S/N=60, which are 
already shown in Figure \ref{st_snr}.

\begin{figure*}
\centering
\includegraphics[angle=0.0,scale=0.8,origin=lb]{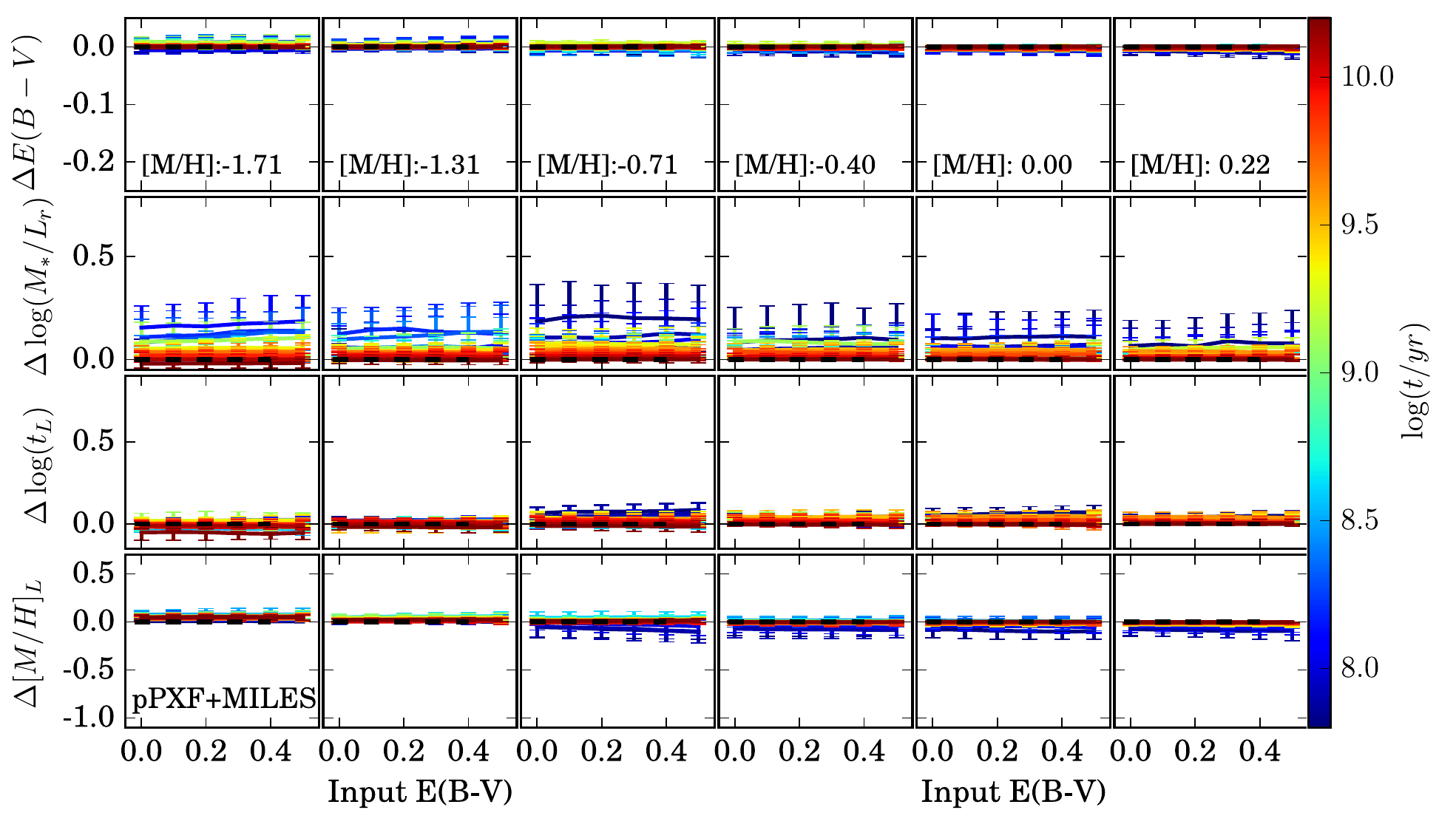}
\caption{
The pPXF fitting of mock spectra generated by single SSP varies in the stellar age and 
metallicity parameter spaces with S/N=30. Lines in different panels are the same as shown in Figure \ref{st_single_ssp}.
}
\label{ppxf_1ssp_snr30}
\end{figure*}

The pPXF fitting results show increased bias and scatter with decreasing S/N. Spectral fitting at S/N=30
can lead to $\sim 2$ times higher parameter biases and scatters than those at S/N=60 (Figure \ref{ppxf_1ssp_snr30}).

\end{document}